\documentclass[acmsmall,nonacm]{acmart}

\AtBeginDocument{%
  }

\setcopyright{acmlicensed}
\copyrightyear{2018}
\acmYear{2018}
\acmDOI{XXXXXXX.XXXXXXX}

\acmConference[Conference acronym 'XX]{Make sure to enter the correct
  conference title from your rights confirmation emai}{June 03--05,
  2018}{Woodstock, NY}
\acmISBN{978-1-4503-XXXX-X/18/06}

\ccsdesc[500]{Human-centered computing~Empirical studies in collaborative and social computing}
\ccsdesc[300]{Human-centered computing~Social network analysis}
\ccsdesc[300]{Information systems~Social networking sites}

\usepackage{graphicx} 
\usepackage{algorithm}
\usepackage{algorithmic}

\usepackage{xcolor}

\usepackage{listings} 

\newcommand{\galen}[1]{}
\newcommand{\tim}[1]{}
\newcommand{\amy}[1]{}
\newcommand{\new}[1]{#1}

\usepackage{xspace}
\newcommand{\hide}[1]{}
\newcommand{\xhdr}[1]{\vspace{1.7mm}\noindent{{\bf #1.}}} 

\newcommand{\eg}{\textit{e.g.,}\xspace}

\newcommand{\fig}{Figure\xspace}
\newcommand{\sect}{\S}
\newcommand{\etal}{\textit{et al.}\xspace}

\begin{document}

\title[Perceptions of Moderators as a Large-Scale Measure of Online Community Governance]{Perceptions of Moderators as a Large-Scale Measure of Online Community Governance}

\author{Galen Weld}
\affiliation{
   \institution{University of Washington}
   \city{Seattle}
   \state{Washington}
   \country{USA}
   }
\email{gweld@cs.washington.edu}

\author{Leon Leibmann}
\affiliation{
   \institution{University of Washington}
   \city{Seattle}
   \state{Washington}
   \country{USA}
   }
\email{lleibm@cs.washington.edu}

\author{Amy X. Zhang}
\affiliation{
   \institution{University of Washington}
   \city{Seattle}
   \state{Washington}
   \country{USA}
   }
\email{axz@cs.washington.edu}

\author{Tim Althoff}
\affiliation{
   \institution{University of Washington}
   \city{Seattle}
   \state{Washington}
   \country{USA}
   }
\email{althoff@cs.washington.edu}


\renewcommand{\shortauthors}{Weld \etal}

\begin{abstract}
Millions of online communities are governed by volunteer moderators, who shape their communities by setting and enforcing rules, recruiting additional moderators, and participating in the community themselves. These moderators must regularly make decisions about how to  govern, yet measuring the `success' of governance is complex and nuanced, making it challenging to determine what governance strategies are most successful. Furthermore, prior work has shown that communities have differing values, suggesting that  `one-size-fits-all' approaches to governance are unlikely to serve all communities well. In this work, we assess governance practices on reddit by classifying the sentiment of community members’ public discussion of their own moderators. We label 1.89 million posts and comments made on reddit over an 18 month period. We relate these perceptions to characteristics of community governance, and to different actions that community moderators take. We identify types of communities where moderators are perceived particularly positively and negatively, and highlight promising strategies for moderator teams. Amongst other findings, we show that strict rule enforcement is linked to more favorable perceptions of moderators of communities dedicated to certain topics, such as news communities, than others. We investigate what kinds of moderators are associated with improved community perceptions upon their addition to a mod team, and find that moderators who are active community members before and during their mod tenures are seen more favorably. We make our models, anonymized datasets, and code public.
\end{abstract}


\maketitle

\section{Introduction}\label{sec:intro}
Millions of online communities rely on volunteer moderators to carry out governance. These moderators perform essential services that shape their communities, including setting and enforcing rules \cite{fiesler_2018_reddit_rules, chandrasekharan_internets_2018, Seering2019ModeratorEA}, 
communicating their actions to the community \cite{jhaver_2019_removal_reasons, zhang_policykit_2020}, and recruiting new moderators. 
Some form of governance provided by moderators is essential to the healthy functioning of almost all online communities \cite{kraut2012building, matias_civic_2019}.

\new{
Mod teams must determine how many moderators to have, how to recruit new mods, what rules to set, and answer many more questions.
With so many decisions to make, it's challenging to determine what governance practices are most effective at ensuring high quality outcomes for online communities. There are so many different communities, with different sizes and topics, that there is no `one-size-fits-all' solution~\cite{weld_values_taxonomy_2021, weld_2022_survey_icwsm}. However, the variety of existing communities presents an opportunity: if we can develop a method to assess the success of communities' governance practices, we can leverage the natural diversity of these practices to identify the most promising strategies for moderators. Doing so is challenging. Survey methods are effective, but expensive to deploy and cannot enable the use of historic data. Notions of `success' are multifaceted, and while classifiers have been used to detect specific harms such as misinformation~\cite{weld_2021_news_sharing} or specific aspects of governance such as rules enforcement~\cite{chandrasekharan_internets_2018}, current methods are unable to quantify broader notions of successful governance.
}

In this work, we measure online community governance by examining how community members themselves perceive their moderators. We develop a method to classify how community members discuss their moderators, publicly, within their communities (\sect\ref{sec:methods}). We use this method to gather and characterize community members' perceptions of moderators at a massive scale, enabling the largest study-to-date of governance practices that we are aware of. We label 1.89 million posts and comments from 8,477 unique subreddits across an 18-month period from January 2020-June 2021, and relate these data to different kinds of online communities, and to different actions that community mod teams can take.

Our analyses address two key research questions:
\begin{itemize}
    \setlength{\itemindent}{0mm}
    \setlength{\itemsep}{0em}
    \item [\textbf{RQ1}] How are moderators of different communities perceived differently by their communities? (\sect\ref{sec:rq1})
    \item [\textbf{RQ2}] What can moderators do to improve how they are perceived by their community? (\sect\ref{sec:rq2})
\end{itemize}

\noindent
We find that community members' perceptions of their mods vary substantially from community to community (\sect\ref{sec:rq1}). Hobby communities have the most positive perceptions of their mods, with meme and news communities having the poorest. Communities that perceive themselves as having high quality content, and being trustworthy, engaged, inclusive, and safe all perceive their mod teams more positively than communities at consider themselves low quality, untrustworthy, disengaged, uninclusive, or unsafe. Community size is also a major differentiator---tiny communities (1-10 posts+comments/day) use $6.1\times$ as much positive language to describe their mods as huge communities (>10k posts+comments/day).

\new{
We identify actionable strategies for community moderators, using IPTW and DID causal inference methods to control for confounding factors including the topic and size of communities (\sect\ref{sec:iptw_did}).
}
We show that communities with fewer than 5 daily posts+comments per moderator use $2.5\times$ more positive language to describe their mods as communities with $20\times$ more posts+comments per mod. Yet our findings do not suggest that more strict moderation improves perception of moderators; in fact, for most types of communities, a 3 percentage point increase in removed content is associated with a 9pp increase in negative language used to describe mods, although news-sharing communities are a notable exception to this trend (\sect\ref{subsec:modteam_size_actions}). We \new{measure the impact of} adding different types of moderators to a community, and find that moderators who are also active community members before and during their tenure as moderators are most strongly associated with improved perceptions of mods (\sect\ref{subsec:mod_engagement}).

We discuss the limitations of our methods, and identify key areas for future work (\sect\ref{sec:discussion}). We make our models, datasets, and code public\footnote{Please contact the authors for additional information.}
to enable further research on this important topic.
\section{Related Work}\label{sec:related}
\xhdr{Measuring Community Governance}
Comprehensively measuring community governance is challenging, as data access to moderator actions is often restricted, except for some types of communities, such as Wikipedia \cite{Panciera2009WikipediansAB} and some gaming servers \cite{frey_2019_minecraft_gov}. On reddit and most popular social media platforms, however access to moderation actions is difficult to obtain, and requires separate permission from the mods of every community to be studied  \cite{li_2022_modlogs, li_measuring_2022}. In contrast, by utilizing public discourse around moderators, our method works for every community whose content is public. 

Because of these challenges, many researchers turn to surveys to study governance, sometimes qualitatively \cite{matias_civic_2019}, and sometimes at a larger scale, for instance to quantify harassment of mods \cite{almerekhi_2020_mod_harassment_modeling} or moderator recruiting practices \cite{seering_2022_twitch_moderator_recruiting}.

\xhdr{Governance Through the Lens of Rules}
Rules are often used as a lens to study governance. Large-scale analyses of rules have been used to characterize governance at the platform level on reddit~\cite{reddy_2023_evolution_rules, fiesler_2018_reddit_rules} and on Wikipedia \cite{Hwang2022_wikipedia_rules}, however it is challenging to infer much about governance in specific communities, as rules constitute only a small portion of governance activity. Surveys of community members' attitudes towards rules have been used to understand members' attitudes towards governance more broadly \cite{koshy_2023_user_mod_alignment}, however this method is challenging to scale beyond a single community. Our method can be applied to thousands of communities.

\xhdr{Moderation Strategies on Reddit}
Rules and their enforcement can also be evaluated in terms of their impact on the community as a whole \cite{jhaver_2019_removal_reasons, srinivasan_2019_content_removal_cmv, jhaver2023bystanders} and their embodiment of communitywide norms \cite{chandrasekharan_internets_2018}.
Beyond rules and their enforcement, researchers have also studied other moderator strategies on reddit, including platform level decisions such as community bans and quarantines \cite{Chandrasekharan2020Quarantined, Ribeiro2020_migration}, and community level actions such as user bans \cite{Thomas_2021_bans_behavior} and stickied posts \cite{Matias_2019_Preventing}. Our work quantifies many of these strategies, including content removal, mod interactions with the community, and associates them with positive and negative moderation discourse.

\xhdr{Measuring Outcomes in Online Communities}
Researchers have studied the range of values that community members hold for their communities \cite{weld_values_taxonomy_2021}, yet it is challenging to accurately predict these values automatically \cite{weld_2022_survey_icwsm}. Large-scale embeddings can be used to understand community culture \cite{Waller_2020_embeddings}, yet not governance directly. Longitudinal work has examined how communities fare with massive growth \cite{Lin_2017_Better_smaller} and the lifecycles of their members\cite{dnm_2013_no_country}, while some research focuses on smaller scale outcomes at the conversation level \cite{Bao_2021_prosocial}. Our work complements this literature with a large scale method for studying governance-specific outcomes.
\section{Methods}\label{sec:methods}

Reddit is a popular social media platform that is frequently studied in the computational social sciences~\cite{Proferes_2021_reddit_research_overview}. Reddit is composed thousands of communities, known as subreddits, each with its own community norms, rules, membership, and moderators.
These attributes, along with the the fact that almost all content on reddit is publicly available, make it an ideal platform for studying community governance at a large scale.

In this work we make extensive use of publicly available reddit data from the Pushshift~\cite{baumgartner_pushshift_2020}. From these data, we detect and classify posts and comments which discuss moderators to produce our dataset of mod discourse (\sect\ref{sec:outcomes}). 
In addition, we collect supplementary data about moderators (\sect\ref{sec:timelines}) and communities (\sect\ref{sec:topic_classifier}).

\subsection{Computing Moderator Timelines}\label{sec:timelines}
An understanding of \textit{who} moderates \textit{which} subreddits \textit{when} is critical to our analyses. We reconstruct timelines of the 10,000 largest subreddit's moderators using  reddit's API and snapshots from the Wayback Machine, a web archiving service provided by the Internet Archive~\cite{internet_archive}. 
We start by scraping current moderator info pages for the 10,000 largest subreddits on the platform, directly from reddit. Each such page contains a list of the current moderators, in order of seniority, along with the exact timestamp that each moderator was added as a moderator to the subreddit. We then use the publicly accessible API from the Internet Archive’s Wayback Machine to scrape every archived copy of every subreddits’ moderator info page going back to 2010. We scrape 30,302 historical archived copies of moderator info pages, which, combined with the 10,000 present-day copies (one per subreddit), give us detailed timelines of who moderated each subreddit when, and when they started as a moderator. Due to the functionality of the Wayback Machine, we have higher temporal resolution (more archived snapshots) for larger and more popular subreddits, yet, because each moderator info page encodes the exact timestamp for each moderator's start date, having fewer snapshots results only decreased accuracy of when moderators resign their posts, not when they are appointed — which is the primary focus of many of our analyses. Since the exact end time of a moderator’s tenure can only be inferred from examining when they were removed from the list of moderators for each subreddit, we adopt a conservative strategy which deliberately underestimates the length of moderators’ tenures: we consider the end date of their tenure as that of the last snapshot for which they were still listed as a moderator. Our method only misses moderators whose appointment, entire tenure, and resignation all occur within the `gap' between snapshots\footnote{We believe this is rare, as the average mod tenure we can detect is longer than four years, and the mean gap between snapshots is 56 days.}. Note that to preserve moderator privacy, we do not make our moderator timelines public.

\subsection{Community Topic and Health Measures}\label{sec:topic_classifier}
Understanding a community's topic and health is critical for understanding that community's perceptions of their moderators. We used a few-shot GPT-4-based classifier to classify the topic of every subreddit included in our analyses, based on their names, into six topical categories taken from existing work: Discussion communities, hobby communities, meme communities, news communities, and video/picture-sharing communities~\cite{weld_2022_survey_icwsm}. The prompt used for this class is given in Appendix~\ref{app:topic_prompt}. We used the manually-labeled dataset of 123 subreddits from \citet{weld_2022_survey_icwsm} for our few-shot examples, as well as to evaluate the performance of the classifier, which has 86.1\% accuracy on the test set, with a macro-average $F_1$ score of 0.858. To understand communities' health, we leverage a recent survey of members of 2,151 different subreddits. Survey participants were asked to rate the current state of their community on an 11-point Likert scale with regards to nine aspects of community health such as the quality of content, the trustworthiness of the community, and the safety of the community \cite{weld_2022_survey_icwsm}. We average across all survey responses for each subreddit to compute an overall subreddit score for each value. Our analyses of differences between communities with different topics and health aspects are in \sect\ref{sec:rq1}.


\subsection{Detecting and Classifying Community Members' Perceptions of Moderators}\label{sec:outcomes}
In this work, we quantify community members' publicly stated \emph{perceptions} of the quality of moderation by automatically detecting and classifying posts and comments which discuss moderators. Our detection and classification pipeline consists of three stages: (1) a prefilter step which uses regular expressions to identify content written by non-moderators which include the words `mod(s),' `moderator(s);' (2) a detection step, which detects content discussing moderators (differentiating them from those which make other use of `mod', \eg `video game mods'); and (3) a classification step, which classifies the sentiment of this content into positive, neutral, negative, and exclude classes. We only include content written by non-moderators as the goal of our work us to examine how non-moderators perceive their mod teams, not how mods discuss themselves in public.
We apply our pipeline, which we make public, to all reddit posts and comments made from Jan. 2020 to June 2021.

\begin{figure}[t]
    \centering
    \includegraphics[width=.8\columnwidth,trim={1mm 94mm 176mm 0mm},clip]{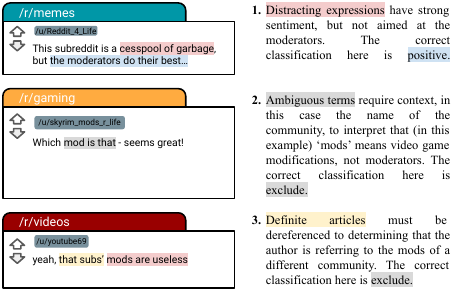}
    \caption{
    Determining the sentiment with regards to the moderators of comments can be very challenging.
    }
    \label{fig:sentiment_example}
\end{figure}

\xhdr{Prefilter Step Details (Step 1)}
The prefilter step efficiently identifies posts and comments which discuss moderators. We use a regular expression-based filter to find all posts and comments which include the words `mod(s),' `moderator(s), and use our moderator tenure timelines to exclude posts and comments written by moderators during their appointment, as our goal is to measure how non-moderator community members discuss moderators, not how moderators discuss themselves.

\xhdr{Detection Step Details (Step 2)}
On reddit, the term `mods' often refers to `moderators,' but often is used as shorthand for `modifications,' as in `video game mods' or `car mods.'
To differentiate posts and comments which discuss moderators from those which use `mod(s)' in other senses, we fine-tuned a RoBERTa-based binary classifier \cite{Liu_2019_RoBERTa} using a manually-labeled dataset of 1,155 posts and comments, randomly sampled from the prefilter step results and further divided into a training set of 655 and a test set of 500 posts and comments. RoBERTa was chosen for its high-performance, and its relative ease of training and minimal compute requirements. To improve performance and provide additional context to the model, we input not only the body of comments (or title and selftext of posts), but also the name of the subreddit the comment/post was made in, as well as the parent comment that the comment being classified was in reply to, when applicable. Our fine-tuned detector model has a precision of 0.82 and a recall of 0.94 on the test set, with an $F_1$ score of 0.88. 

\xhdr{Classification Step Details (Step 3)}
The classification step classifies posts and comments based on their sentiment \emph{with regards to the community moderators.} A comment with an overall-negative sentiment may have a positive sentiment with regards to the moderators, and vice versa (\fig\ref{fig:sentiment_example}). To label posts and comments for the classification task, two annotators worked together to iteratively refine a codebook (Appendix \ref{app:codebook}), then independently labeled 200 randomly sampled posts/comments. The annotators had `almost perfect' inter-annotator reliability (0.85 Fleiss' kappa) \cite{landis_koch_1977_measurement}. To produce a `gold standard' test set, the same two annotators independently labeled a random sample of 500 posts/comments, then discussed their disagreements until consensus was reached. A single annotator then labeled an additional 734 posts/comments for use as a training set, of which 484 were randomly sampled, and 250 were sampled based on their proximity to the decision boundary of a simpler RoBERTa model trained for the classification task.

\begin{table}[t]
    \centering
\begin{tabular}{l|rrrrr}
                 & Human           & Random & VADER  &  GPT-4  & Our Model       \\ \hline
Positive $F_1$   & \textit{0.85}   & 0.14   & 0.30   & 0.70   & \textbf{0.71}    \\
Neutral  $F_1$   & \textit{0.89}   & 0.50   & 0.34   & 0.71   & \textbf{0.73}    \\
Negative $F_1$   & \textit{0.84}   & 0.13   & 0.34   & 0.61   & \textbf{0.71}    \\
Exclude  $F_1$   & \textit{0.98}   & 0.21   & 0.00   & 0.66   & \textbf{0.73}    \\ \hline
Test Set Acc.    & \textit{89.0\%} & 27.0\% & 30.6\% & 66.2\% & \textbf{72.4\%}        
\end{tabular}
    \caption{ \normalfont
    \textbf{Our Classification Step model, a LLaMA 2 model fine tuned with QLoRA \cite{touvron2023llama2, dettmers2023QLoRA}, exceeds the performance of a retrieval-based few-shot classifier using GPT-4.} Our model is also more scalable (it can be deployed locally), more affordable, and more reliable (it is not subject to prompt filtering), than GPT-4. This table compares $F_1$ scores for expert human labelers, retrieval-based few-shot GPT-4, and our model, alongside an empirical class distribution random baseline and a VADER-based classifier \cite{hutto_2014_VADER}.}
    \label{tab:model_performance}
    \vspace{-8mm}
\end{table}

As the purpose of this method is to identify perceived moderation quality, we must use care to identify the moderators that posts/comments are discussing. On reddit, community members occasionally discuss moderators of \emph{other} communities, \eg a member of /r/gaming praising the moderators of a specific Discord server, or a member of /r/nfl complaining about the moderators of /r/seahawks. To ensure correct attribution, we decided to limit our analyses to community members talking about their \emph{own communities'} moderators (\eg discussion of the /r/cats moderators taking place on /r/cats).
For this, we specifically trained our model to identify posts and comments which discuss the moderators of other communities, along with content that erroneously passed the detection step. We trained on this `exclude' class in addition to those used for downstream analyses: positive, neutral, and negative.
\footnote{Additionally, we attempted to classify specific complaints about moderation: excessive moderation, insufficient moderation, and biased moderation, but found classifier performance not to be amenable for high-confidence downstream analyses.}

To further enhance the performance of our model, we performed data augmentation using a retrieval-based few-shot classifier built with GPT-4 \cite{achiam_2023_GPT4}. We used this classifier to label an additional sample of 10,000 posts and comments, which were combined with our manually-labeled training set to fine-tune our final classification model, a 13-billion parameter LLaMA 2 model \cite{touvron2023llama2} fine-tuned with QLoRA \cite{dettmers2023QLoRA}, using the prompt shown in Appendix~\ref{app:sentiment_prompt}. LLaMA 2 and QLoRA were selected for their very high performance while still being feasible to fine-tune and deploy on a massive dataset. Our model was fine-tuned on an internal university HPC cluster with $2\times$ NVIDIA a40 GPUs, which took $\approx 13$ hours. Our final model exceeds the performance of GPT-4 on all classes (Table~\ref{tab:model_performance}). Applying the finalized Classification Step to the results from the Detection Step left us with a labeled dataset from 8,477 communities of 196,231 posts and 1,694,551 comments which discuss moderators: 175,296 with positive sentiment, 968,235 with neutral sentiment, and 747,251 with negative sentiment, and we make an anonymized version of this dataset public.


Accurately classifying the sentiment with regards to the moderators of a post or comment is an extremely challenging task that is often heavily reliant on context and background knowledge of the community (\fig\ref{fig:sentiment_example}). As such, off-the-shelf sentiment classifiers, such as VADER, perform very poorly (Table~\ref{tab:model_performance}). Even general purpose LLMs such as GPT-4, which obtain state-of-the-art results on standard sentiment analysis benchmarks \cite{kheiri_2023_SentimentGPT}, perform worse than our fine-tuned LLaMA2 QLoRA model, even when prompted using a retrieval-based few-shot in-context learning method.

\xhdr{Interpreting Mod Discourse} 
Finally, to enable downstream analyses, we define several aggregate values for each community, used in \sect\ref{sec:rq1} \& \sect\ref{sec:rq2}:
The \emph{Amount of Mod Discourse} for a given community is the fraction of all posts and comments discussing mods in that community, regardless of their sentiment. The \emph{\new{Composition} of Mod Discourse} with positive/negative sentiment is the \new{percent} of all posts and comments discussing mods in a given community that have positive/negative sentiment. To ensure a meaningful amount of data, we exclude all communities that did not have at least one positive, neutral, and negative post or comment discussing moderators during our 18-month analysis period. Finally, we exclude nine communities explicitly devoted to discussing moderators, as it is infeasible to differentiate discussion of those communities' moderators from other moderators. This filtering leaves us with 5,282 subreddits used in downstream analyses. 


\subsection{Validating our Measures of Perceptions of Moderators}\label{sec:outcome_validation}
Our methods for detecting and classifying community members' perceptions of moderators (\sect\ref{sec:outcomes}) measure when community members publicly post or comment in their own communities to discuss moderators. For ethical as well as practical reasons, we cannot measure when community members discuss their moderators \emph{privately}. However, there are two other instances where community members may attempt to discuss their moderators publicly but would not be included in our analyses. In this section, we show that these two cases are very rare relative to the public moderator discourse that we \emph{do} measure.

For purposes of computational efficiency, our initial prefiltering step only includes posts and comments which use the phrase `mod(s)' or `moderator(s)' for subsequent classification. Occasionally, however, community members may discuss their moderators by mentioning their usernames specifically. For example, instead of writing `the mods of this subreddit are unfair,' which would be captured by our pipeline, a community member might write `/u/exampleModerator is unfair.' To assess the frequency of this form of moderator discourse, for a sample period of one month (January 2018, the first month of our analysis period), we compute how frequently the usernames of \emph{any active moderator} is mentioned directly in their own subreddit, in addition to the frequency of uses of `mod(s)' or `moderator(s).' We find that such specific moderator mentions are very rare, with the average subreddit having 64.11 generic uses of `mod(s)' or `moderator(s)' for every time a moderator is mentioned specifically by their username. We assessed the impact that including these specific mentions would have on downstream analyses, and concluded qualitatively that they did not make a substantial difference in our results. Thus, for purposes of computational efficiency, we chose to exclude specific moderator mentions from our moderator discourse classification pipeline.

An additional potential source of bias in our analyses is that of removed posts and comments. On reddit, moderators are occasionally accused of removing content in their communities that is critical of the moderators~\cite{matias_civic_2019}. As this content could be removed before entering our moderator discourse classification pipeline, it is conceivable that our results undercount moderator discourse in communities where the moderators remove such content. Assessing this source of bias requires knowledge of the content that is removed, which is technically infeasible to collect at a large scale. However, prior researchers have collected content before it is removed by moderators by scraping content from specific subreddits as it is posted, then checking later to see if the content has been removed~\cite{chandrasekharan_internets_2018, Kumar2023WatchYL}. We used the Pushshift API (before it was taken offline) in the same manner, to collect the text of content that was removed from 100 randomly sampled subreddits over a two year period (2017-2018, inclusive). Using this method, we collected 263,657 pieces of removed content from 100 different communities. Of the content that was removed by moderators in this sample, we find that it is extremely rare for moderators to remove content that mentions the moderators: The average community in our sample removes 102 posts/comments which do not mention the moderators for every post or comment removed that does mention the moderators. Once again, we assessed the impact that including removed comments would have on our downstream analyses, and concluded that their inclusion would not make a substantial difference in our results (in addition to be infeasible at the scale of the rest of this work).
\section{How are moderators of different communities perceived differently by their communities?}\label{sec:rq1}
Online communities exist for nearly every conceivable topic, and range in size from just a few members to many millions. In this section, we examine how community members' perceptions of their moderators vary across communities with different topics, different community health metrics, and different sizes.

\begin{figure*}[ht]
    \centering
    \includegraphics[width=\textwidth]{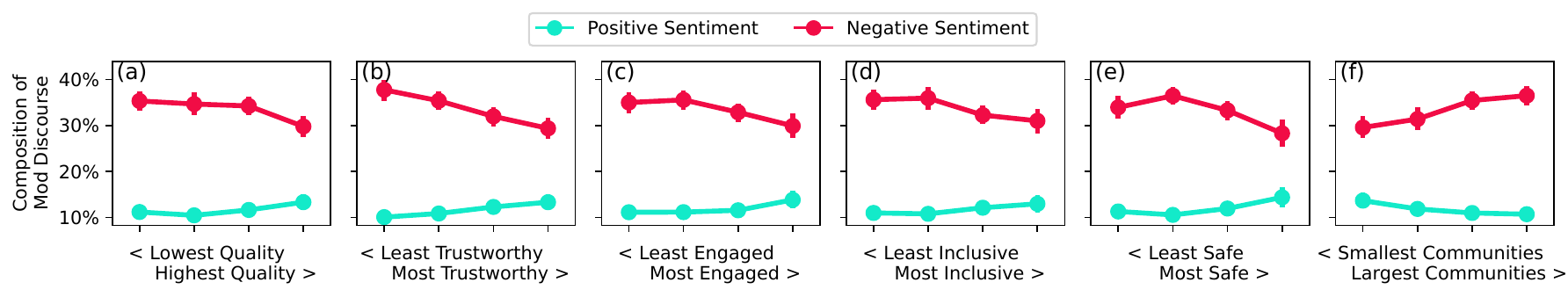}
    \vspace{-9mm}
    \caption{ \normalfont
    \textbf{Communities that consider themselves higher quality (a), more trustworthy (b), more engaged (c), more inclusive (d), and more safe (e) all use more positive and less negative sentiment to describe their moderators.} Here, communities are grouped into quartiles based on their community members' self-reported perceptions of the current state of the community. This effect is most pronounced for communities' self-reported trustworthiness (b), with the top-25\% most trustworthy communities using 34\% more positive and 22\% less negative language to describe their mods. Communities that rate themselves as feeling smaller (f) have a more positive perception of their mods. In this and all other figures, points represent mean estimates alongside bootstrapped 95\% confidence intervals.}
    \label{fig:values}
\end{figure*}

\xhdr{Method}
To quantify a community's health, we divide communities into quartiles based on their community members' responses to a recent survey (\sect\ref{sec:topic_classifier}). Survey responses were collected between May-July 2021, a period overlapping the the end of our mod discourse data time range. For this analysis only, we exclude communities which were not surveyed. Community topic is classified into one of six topical categories using our topic classifier (\sect\ref{sec:topic_classifier}).  We quantify community size as the volume of submitted content: the average number of posts and comments per day over our study period, which we use to group communities with similar size. For each group, we compute the amount of mod discourse as well as the composition of that discourse.

\xhdr{Results}
Different aspects of community health are associated with better and worse perceptions of moderators. The smallest-feeling 25\% of communities, based on member self-reports, use 27\% more positive and 16\% less negative sentiment to discuss their moderators than the largest-feeling 25\% of communities (\fig\ref{fig:values}a). \fig\ref{fig:values}a-e shows that communities which rate themselves as having higher quality content and being more trustworthy, more inclusive, and more safe all use more positive and less negative to discuss their moderators, as well.


\begin{figure*}
    \centering
    \includegraphics[width=\textwidth]{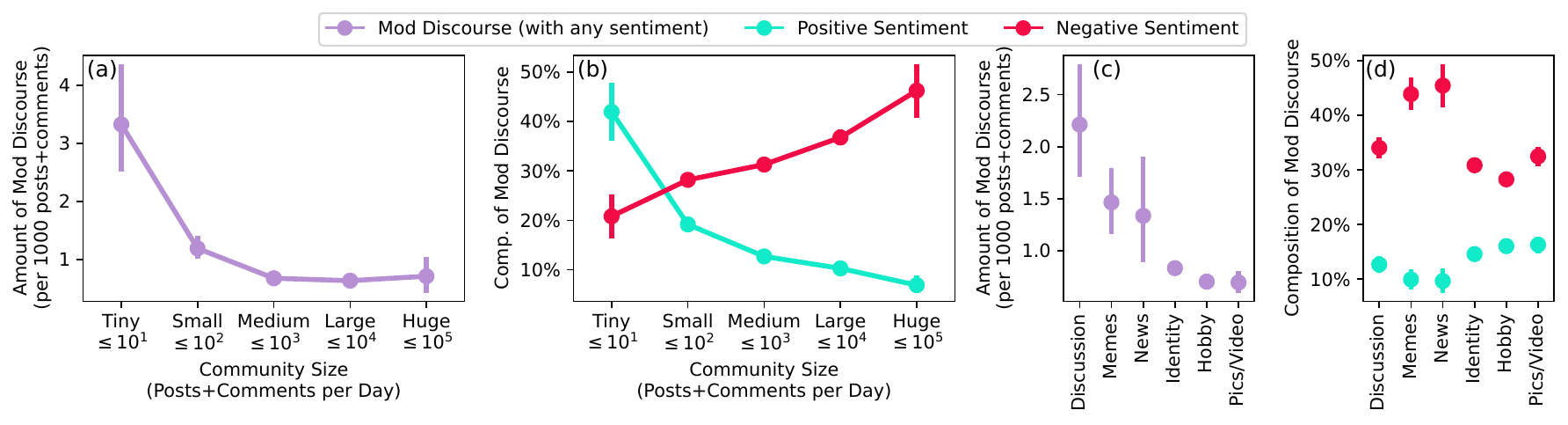}
    \vspace{-8mm}
    \caption{ \normalfont
    \textbf{Perceptions of moderators vary significantly across communities with different sizes (a-b) and topics (c-d).} In general, smaller communities devote a relatively larger proportion of their content to discussing their moderators (a), and smaller communities express more positive and less negative sentiment towards their mods (b). Discussion, meme-sharing, and news communities have proportionally more mod discourse (c), while meme and news-sharing communities exhibit the most negative sentiment towards their moderators (d).
    }
    \label{fig:size_topic_combined}
\end{figure*}

Direct measurement of communities' size allows us to further investigate the relationship between the volume of content submitted to a community, and its members' perceptions of its moderators. 
\new{
In general, smaller communities have more mod discourse (\fig\ref{fig:size_topic_combined}a), with tiny communities with fewer than ten posts+comments per day having $7.6\times$ as many posts and comments discussing moderators as huge communities (those with more than 10k posts and comments per day), relative to the total amount of content. Tiny communities also have $6.0\times$ more positive mod discourse, and $0.46\times$ as much negative discourse, than huge communities.
}


Examining communities of different topics, \new{
discussion communities have the most mod discourse, on average (\fig\ref{fig:size_topic_combined}c).
}
Hobby communities have generally the most positive mod discourse, with meme and news-sharing communities having 46\% more of their mod discourse have negative sentiment than communities of other topics (\fig\ref{fig:size_topic_combined}d).


\xhdr{Implications}
Surveys of redditors have shown that they consider a wide range of factors to be important to the overall `health' of their communities \cite{weld_values_taxonomy_2021}. Here we show that communities that are doing well with regards to factors widely considered to be important, such as safety and quality of content, tend to perceive their moderators more positively. Many of these factors are only indirectly controlled by the moderators, for example quality of content---while moderators can set and enforce rules aimed at improve the quality of content in their communities, quality of content is ultimately a function of the content submitted by community members, not moderators. This can lead to moderators being `blamed' for problems largely outside of their control, which has been previously shown in small-n surveys \cite{matias_civic_2019, jhaver2023decentralizing} and is further supported empirically here, as shown in \fig\ref{fig:values}.

Small communities appear to both discuss their moderators more, as well as use more positive sentiment in their discussions (\fig\ref{fig:size_topic_combined}a,b). Several factors may contribute to this, including that the increased anonymity that comes with participating in a larger community may make people feel more comfortable speaking negatively about the moderators, and that smaller communities are more likely to be newly formed and thus still establishing moderation norms, leading to more mod discourse \cite{hwang_2021_small_communities, seering_2022_twitch_moderator_recruiting}.


\section{Adjusting for Confounders using IPTW \& DID}\label{sec:iptw_did}
While we are ideally interested in causal, actionable insights, our study is a retrospective observational study, and like any other, subject to potential confounding. In \sect\ref{sec:rq1} we identified two important confounders, size and topic, which are correlated with perceptions of moderators (\fig\ref{fig:size_topic_combined}). Next we describe two methods from the causal inference literature which we use to adjust for these and other confounding factors in subsequent analyses.

\xhdr{Inverse Probability of Treatment Weighting (IPTW)}
IPTW is a statistical method to adjust for imbalance in possible confounding factors when making comparisons between different observed groups (\eg a treatment and a control group). We select IPTW over other methods, such as stratification, due to its excellent efficiency~\cite{Austin2015_IPTW_best_practice}. In \sect\ref{sec:rq2}, we compute the probability of treatment (propensity score, $P(Z|\mathbf{X})$) by applying a logistic regression model to community covariates $\mathbf{X}$ including the topic and size of the community. A complete list of covariates for each analysis is given in Appendix~\ref{app:covariates}. We generalize IPTW to non-binary treatments to estimate a dose-response curve using the method from \cite{althoff2016quantifying} in which the treatment is discretized into bins and a different propensity score is computed for each treatment bin in a `one-versus-rest' scheme. For final analyses, each observation is weighted by the inverse of the probability of the treatment it received, such that the weight for observation $i$ is $w_i = \frac{Z_i}{P(Z_i=1|\mathbf{X}_i)} + \frac{1-Z_i}{P(Z_i=0|\mathbf{X}_i)}$, where $Z_i$ is the treatment received by observation $i$ (0 for control, 1 for treated), and $\mathbf{X}_i$ is a vector of the covariates.
In analyses where we use IPTW, we also include the non-IPTW adjusted estimate in figures with a gray color. In these analyses, reweighting does not dramatically impact our findings, suggesting that the impact of these confounders is moderate at best.

An important validity check for IPTW is to assess the balance of covariates for each treatment group after weighting as applied~\cite{Austin2015_IPTW_best_practice}. 
Two groups are often considered `balanced' or `indistinguishable` if all covariates are within a standardized mean difference (SMD) of 0.25 standard deviations~\cite{althoff2016quantifying}.
For each treatment group, we compute the difference between each covariate's weighted mean value and the \emph{reference distribution}, consisting of the entire population. We compute the standardized mean difference (SMD) by normalizing the difference in means by the standard deviation of the values of the reference distribution.
Our reweighting method achieves balance for every covariate substantially related to the treatment.
SMDs after weighting are given for each covariate and each analysis in Appendix~\ref{app:covariates}.

\xhdr{Difference in Difference Analyses (DID)}
\new{
Difference in difference analyses enable the estimation of the impact of an intervention by comparing the outcome \emph{before vs. after} the intervention was applied to the treated group, and comparing this before \emph{vs.} after difference to an untreated control group.
For our analyses of the impact of moderators' community engagement (\sect\ref{subsec:mod_engagement}), we use modified DID analyses with multiple time periods~\cite{callaway_2021_diff_in_diff} to compare the difference in communities' receptions of new moderators who are engaged with the community (the `treatment') \emph{vs.} those who are not (the `control'). Two four week long periods immediately preceding and following each moderators' appointment were used to compute the first-order difference.
Since mod discourse naturally varies over time, averaging over many DIDs from many new moderator appointments serves to reduce confounding background temporal trends by comparing the difference to an untreated control group.
}

\section{What can moderators do to improve how they are perceived by their community?}\label{sec:rq2}
Moderators have a great deal of autonomy to run their communities as they see fit, including enforcing rules by removing content and growing the mod team by recruiting or appointing new moderators. Moderators also may (or may not) participate in the community as `regular community members' in addition to their mod duties. In this section, we \new{identify promising suggestions for moderators by comparing communities whose moderators have different workloads, rule enforcement strategies, and degrees of community engagement.}

\begin{figure}[t]
    \centering
    \includegraphics[width=.6\columnwidth]{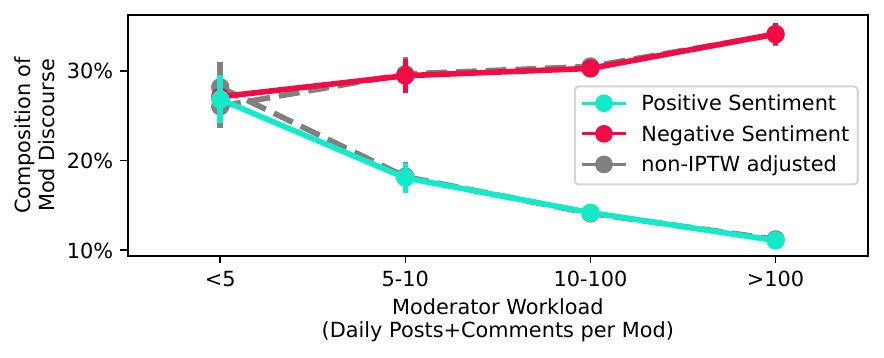}
    \caption{\normalfont
    \textbf{Moderators in communities with lower workloads are perceived more positively and less negatively than moderators in communities with high workloads.} Communities with lower moderator workloads (more moderators relative to the amount of content submitted) tend to have more more positive sentiment in their discussion of the moderators, and less negative sentiment. Communities with fewer than five posts and comments per mod per day use $2.5\times$ as much positive sentiment in their mod discourse compared to communities with more than 100 posts and comments per mod per day.
    }
    \label{fig:num_mods}
    \vspace{-3mm}
\end{figure}

\subsection{Content Removal and Moderator Workload}\label{subsec:modteam_size_actions}
\xhdr{Method}
We can identify content removed by moderators in each community by counting the occurrences of `[removed]' posts and comments within each community. By dividing by the total amount of content submitted to that community, we can compute the total percentage of content removed by mods. Using our mod timelines (\sect\ref{sec:timelines}), we can compute the total number of moderator-tenures in any given subreddit over the course of our analysis time period. We also approximate `workload' of each mod by dividing the total amount of content submitted to each community by the number of mods available to review that content; while in reality it is unlikely that all moderators share the work of reviewing content evenly \cite{li_2022_modlogs}; this metric nonetheless helps us understand the ratio of moderators to content within each community. If workload was shared unevenly, this would only make the workload even higher for a subset of moderators, leading to at worst to overly conservative estimates (higher workload to be addressed by fewer moderators).

\xhdr{Results}
We find that communities with higher mod workloads have more negative mod discourse (\fig\ref{fig:num_mods}). Communities with fewer than five pieces of daily content for each mod (8.8\% of all communities) have approximately equal amounts of mod discourse with positive and negative sentiment, a rarity on a platform where mods are far more commonly discussed negatively. In communities with higher mod workload, the composition of mod discourse is much less positive, with communities with more than 100 posts and comments per mod per day using positive sentiment to describe their mods only $0.39\times$ as often as the 8.8\% of communities with the lowest mod workload.

\begin{figure}[t]
    \centering
    \includegraphics[width=.6\columnwidth]{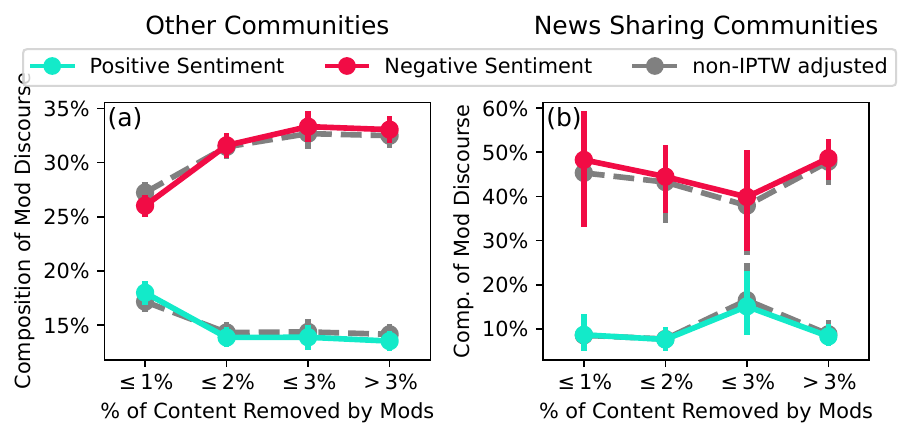}
    \caption{\normalfont \textbf{For \new{most topics, communities where moderators remove more content exhibit \textit{more} negative sentiment}} (a). News communities, however, buck this trend, with the fraction of mod discourse with negative sentiment \new{11 percentage poitns \textit{lower} in news communities whose mods remove less than $1\%$ of content compared to} communities whose mods remove 2\% - 3\% of content (b).}
    \label{fig:removed_content_stratified_news}
\end{figure}

Generally, we find that the composition of mod discourse is more negative in communities with more removed content (Supplementary \fig~\ref{fig:removed_content}). However, these trends vary depending on the topic of the community, with news communities in particular showing the opposite trend (\fig\ref{fig:removed_content_stratified_news}).\footnote{Similar figures for each of the six community topics are included in Supplementary \fig\ref{fig:removed_content_stratified_all_topics}.} Amongst news communities that remove $\leq1\%$ of their content, 48\% of mod discourse has negative sentiment, which \textit{drops} 11pp to 39\% for news communities whose mods remove between 2\% and 3\% of their content. For the same amounts of removed content for non-news communities, the fraction of mod discourse with negative sentiment \textit{increases} 6pp, from 26\% to 33\%.

\xhdr{Implications}
Our results suggest that adding additional moderators to a community (and therefore reducing the effective moderator workload) may improve community members' perceptions of their mods (\fig\ref{fig:num_mods}). 
In \sect\ref{subsec:mod_engagement} we report on additional longitudinal evidence that recruiting additional moderators can have a positive effect.
However, simply increasing the amount of content which is removed does not, in general, appear to be associated with more positive moderator discourse sentiment---in fact, the opposite appears to be the case for communities that are not focused on sharing news (\fig\ref{fig:removed_content_stratified_news}). Taken together, these results imply that there are topic-specific nuances to content removal, and that mods should use care when deciding how strictly to enforce rules, and how much content to remove.
Our results suggest that certain topics are more amenable to stricter rule enforcement than others.

\subsection{Community Engagement}\label{subsec:mod_engagement}
\xhdr{Method}
Some moderators are actively engaged with community, soliciting feedback from non-moderators, updating community members on moderation-related news, and contributing to regular community content in addition to their official moderator duties. Other moderators are far less visible, opting to remove content and change rules without participating in the community more broadly. Furthermore, many moderators also participate in other communities beyond just the one(s) they moderate. Using moderators' public posts and comments, for each moderator-appointment to a community, we compute the number of posts and comments they made in that community and in other communities before and during their tenure as a moderator.
\new{
We use DID analyses (\sect\ref{sec:iptw_did}) to estimate the impact of appointing new mods with different degrees of community engagement (or lack thereof).
}

\begin{figure}[t]
    \centering
    \includegraphics[width=.6\columnwidth]{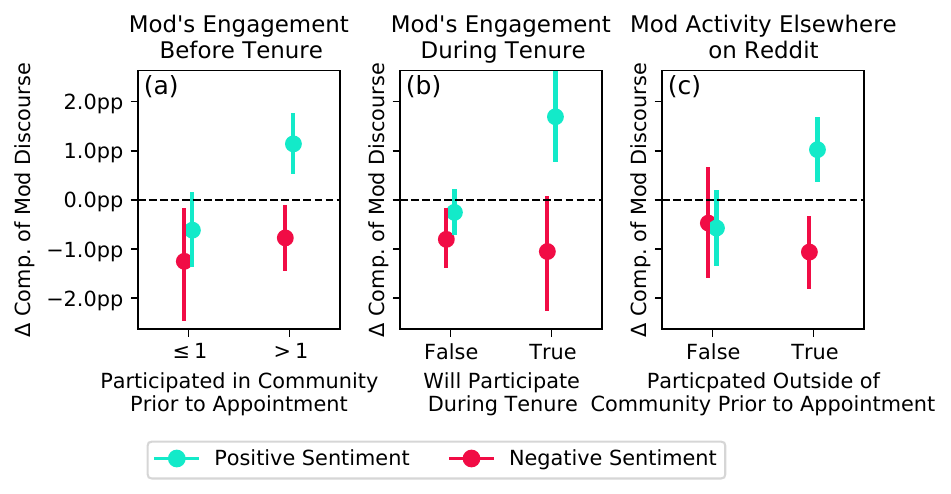}
    \caption{\normalfont
    \textbf{Newly appointed mods are associated with a greater improvement in mod perceptions if they are engaged in the community and elsewhere on reddit before their tenure, and if they are engaged during their tenure (a-c).}
   Adding a moderator who already has or will engage with the community \new{is associated with} in an increase in the fraction of mod discourse in the community with positive sentiment (aqua), and that increase is 32.5\% larger when adding a mod who will engage with the community going forward (b) than for one who already has (a).  Adding a moderator who is an active member of communities other than the one they are becoming a mod of \new{is associated with} an increase in positive, and a decrease in negative, sentiment in mod discourse (c).}
    \label{fig:engagement}
    \vspace{-5mm}
\end{figure}

\xhdr{Results}
We find that, regardless of the moderators' previous engagement with the community, adding a new moderator to a community \new{is associated with a} decrease in mod discourse with negative sentiment (\fig\ref{fig:engagement}). However, not all moderators have the same impact. The most positively impactful moderators are those who are engaged with the community prior to their appointment (\fig\ref{fig:engagement}a), will continue to engage with the community \emph{during} their tenures (\fig\ref{fig:engagement}b), and were also active in \emph{other} communities prior to their appointment (\fig\ref{fig:engagement}c). New mods who do all three of these things are associated with a 2.5pp increase in positive mod discourse, whereas mods who do none of these things are associated with a 0.5pp \emph{decrease}.


\xhdr{Implications}
Our results suggest that moderators who actively engage with the community both before and during their tenures are most likely to have a positive impact on the community's perception of their mod team, which may be because these moderators are more familiar with the community, and are better mods as a result. Another plausible mechanism for this effect is that non-moderator community members value the transparency and accountability that may stem from increased moderator engagement. Our results also suggest that moderators who are active in other subreddits beyond the one(s) they moderate have a more positive impact on the communities they moderate, perhaps because participating in a broader range of communities makes them more effective moderators \cite{zhu_2014_wikia_membership}. Lastly, our results suggest that of these factors, engagement with the community \textit{during} the moderator's tenure has the largest impact.

\subsection{Moderator Recruiting}\label{subsec:mod_recruiting}
When it's time to grow the mod team, existing moderators have a wide array of options for who to recruit, and how to recruit them, and little guidance, official or otherwise, for how to select new mods. In practice, moderator recruiting falls into two different strategies: \emph{Public Recruiting}, where moderators post publicly \emph{internally} in their own subreddit that they are looking for moderators, and solicit applications, nominations, or hold elections. Sometimes, moderators make use of special \emph{external} moderator-recruiting subreddits, such as /r/needamod, where they can post `job listings' to prospective applicants. By contrast, \emph{Private Recruiting} is the recruiting of new moderators in the absence of any public recruiting activity. Moderators recruited privately are either invited to join directly by an existing mod, make an offer to moderate themselves, or are recruited via other backchannel means.

\begin{figure}[t]
    \centering
    \includegraphics[width=.6\columnwidth,trim={0 3mm 0 2mm},clip]{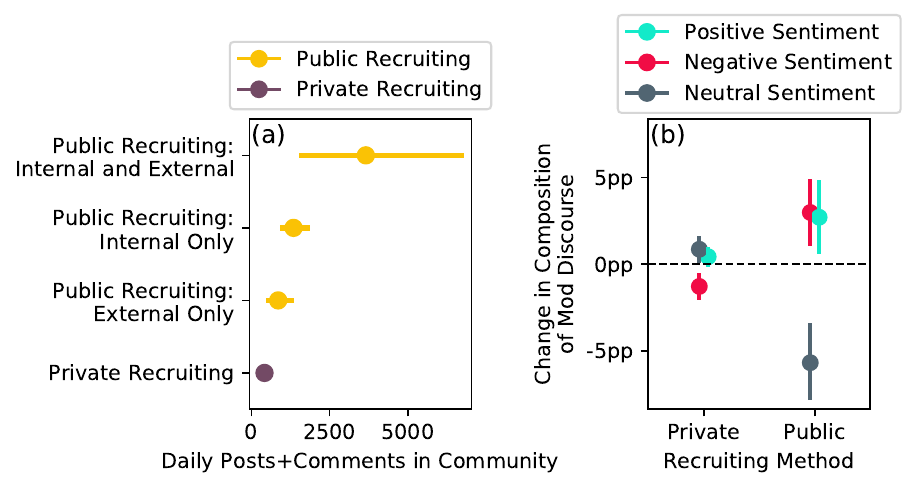}
    \caption{ \normalfont
    \textbf{Communities that recruit moderators publicly are $8.78\times$ larger than the average community which recruits only privately,} in terms of the community's daily volume of content (a). Small communities lean towards private recruiting. (b) Compared to private recruiting, recruiting moderators publicly is polarizing: \new{it is associated with} an increase in \textit{both} positive and negative fractions of mod discourse, and a corresponding decrease in neutral sentiment.}
    \label{fig:recruiting}
    \vspace{-5mm}
\end{figure}

\xhdr{Method}
We identify instances of public recruiting using a regular expression-based search for posts made by sitting moderators which use `recruiting new mods,' `applications open for new mods,' `holding mod elections,' or similar phrases. We also apply a regular expression-based filter to the complete set of posts from /r/needamod to identify which subreddits are recruiting, restricting the results to only posts made by current moderators of the subreddit that is recruiting.
We then take our dataset of moderator timelines (\sect\ref{sec:timelines}) and match moderators who were appointed to a community within 8 weeks of a public recruiting post as having been publicly recruited. Moderators appointed without a recent public recruiting post are considered privately recruited. 

\xhdr{Results}
We find that, while communities of all sizes use public recruiting methods, larger communities are the most likely to make use of public recruiting (\fig\ref{fig:recruiting}a). In terms of its daily volume of content, the average community that uses both Internal and External Public Recruiting is $8.78\times$ larger than the average community that only recruits privately. Examining the impact that adding a single moderator recruited privately vs. publicly has to a community in \fig\ref{fig:recruiting}b, we find that public recruiting appears to be polarizing, with the fractions of both positive \emph{and} negative mod discourse increasing by 2.7pp and 3.0pp, respectively, after a publicly recruited mod is added, on average.

\xhdr{Implications}
Our findings suggest that public moderator recruiting has the potential be a powerful tool to improve community perceptions of moderators when used carefully, and can be harmful when used without regard to the community's preferences. A plausible explanation for the increased polarization resulting from public recruiting is that in some circumstances, the public nature of the recruiting exacerbates existing frustrations with mod teams, for example if due process was not followed during moderator elections, or if a perceived-outside was brought in via external recruiting when community members themselves preferred someone with more experience in the community. More work is needed to assess the differences between different moderator recruiting strategies.

\subsection{Novice and Experienced Mods}\label{subsec:mod_experience}
\xhdr{Method}
When selecting new moderators, existing moderators may be inclined to favor candidates who already have some moderation experience in other communities. Using Moderator Timelines (\sect\ref{sec:timelines}), we can accurately assess how much experience a new moderator has at the time the are appointed, if any. 

\begin{figure}[t]
    \centering
    \includegraphics[width=.6\columnwidth,trim={0 3mm 0 2mm},clip]{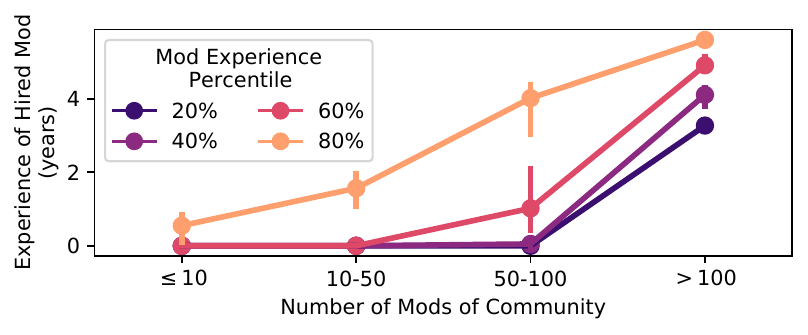}
    \caption{\normalfont \textbf{In contrast to small moderator teams, large teams are much more likely to appoint new moderators who already have moderation experience.} $94\%$ of mods recruited to join mod teams with more than 100 mods have at least 2 years of experience, while 74\% of mods who join small teams with fewer than 10 mods have no previous experience at all.}
    \label{fig:mod_experience}
    \vspace{-5mm}
\end{figure}

\xhdr{Results}
We find that large communities are much more likely to recruit mods who already have moderation experience (\fig\ref{fig:mod_experience}). $74\%$ of mods appointed to communities with fewer than 10 mods are first time mods, while $94\%$ of mods appointed to communities with $>100$ mods have more than 2 years of experience.

\xhdr{Implications}
Large communities with large moderator teams rarely appoint novice moderators, perhaps due to a perception that a large community is not an appropriate place for new moderators to gain experience. As such, it seems that the most common pattern on reddit is for new moderators to start off moderating small communities, and then work their way up to larger ones. This may contribute to a perception, especially amongst non-moderators, that moderators are motivated to increase their number of appointments and to moderate larger and larger communities, potentially biasing the performance of their duties.

\section{Discussion}\label{sec:discussion}
The massive diversity of online communities offers enormous potential to empirically study how to make online communities better. Any such studies, however, require robust and scalable methods to quantify both the outcomes (which communities are doing `better' than others), as well as the independent variables (the aspects of communities that might make them better). Measuring the success of a community's moderators is more challenging than many other aspects of making a community `better,' as unlike, say, misinformation, governance is far less visible in a community, and its success is far less well defined. The primary contribution of our work is the use of community members' own \textit{perceptions} of their moderators, leveraging millions of people's perceptions of good governance, rather than attempting to define good governance ourselves.

We are tremendously excited about the potential synergies this line of work enables. While the findings presented in this paper are impactful by themselves, combining our measure of governance with many other important measures of community outcomes (safety, inclusion, and discussion quality, for example) will enable studies of how to make communities better that are both more comprehensive and more robust than previously possible.

\xhdr{Diversity of Communities' Mod Perceptions}
Our results show that communities' perceptions of their moderators are highly varied (\fig \ref{fig:values} \& \ref{fig:size_topic_combined}) but are associated with other aspects of community health and size. This suggests that not only do different kinds of communities have different norms for their mod discourse, but also that community members' perceptions of overall health of their community, which includes many factors outside of the moderators' direct control, also influences their mod discourse. Researchers and moderators must carefully consider the specific needs of their communities.

\xhdr{Topic Contention Affects Governance}
Our results suggest that the topic of a community impacts the community members' perception of moderators' removal of content and rule enforcement (\fig\ref{fig:removed_content_stratified_news}, Appendix \fig\ref{fig:removed_content_stratified_all_topics}). We find that for news communities, as moderators remove more content, community members' perception of moderators becomes more positive (up until $\approx3\%$ of content is removed), while for other communities, perceptions of moderators are more negative as the mods remove more content. This suggests that communities with certain topics, perhaps more contentious ones, are more appropriate to moderate fairly strictly than others. Future work should specifically examine the impact of community topic on community members' perceptions of content removal.

\xhdr{Community Overestimates of Moderators' Power}
There are many aspects of every community that moderators have limited-to-no control over. For example, our results show that perceptions of moderators are most positive in smaller communities (\fig\ref{fig:values}f). However, although they can set and enforce policies with the goal of growing or shrinking a community, moderators cannot directly control a community's size. Thus, our results support previous research that moderators are frequently `blamed' for problems that they cannot easily resolve~\cite{matias_civic_2019}. Future work could address this tension directly, perhaps by attempting to educate community members on the roles and powers of moderators, as well as the powers that moderators \emph{do not} have.

\xhdr{Mod Recruiting and Engagement}
We examine instances of moderators being added to mod teams to compute the impact that different types of moderators have on their communities. We find that more engaged moderators are associated with an increase in the amount of positive mod discourse, perhaps as a result of an increased sense of transparency and/or accountability for community governance (\fig\ref{fig:engagement}a-c). We also look at differences between moderators whose recruiting was discussed publicly, versus those who were recruited in private. We find that publicly recruited moderators appear to be more controversial, with both negative \emph{and} positive discourse increasing after their appointment (\fig\ref{fig:recruiting}b). This suggests that moderators must use caution when recruiting moderators publicly, and be sure to consider community preferences.

\subsection{Limitations}\label{sec:limitations}
Our work measures community governance through community members' public discussion of moderators. While these signals enable insights about different governance strategies, they do not capture every aspect of the success of a community's governance. What people say publicly does not always reflect their actual beliefs, and even if it did, minimizing community unhappiness is not necessarily the best objective function for community moderators. Although we conclude that excluding content that mentions specific moderators by name and removed content does not substantially impact our results (\sect\ref{sec:outcome_validation}), differences in community members' behavior and values also may still bias our results, as different community may have different norms around mod discourse, and different community members may feel more or less comfortable expressing their opinions publicly. Additional study is needed to ensure that scalable methods for measuring governance reflect the needs of all community members, not just the noisiest.

Even though our fine-tuned sentiment classifier exceeds the performance of state-of-the-art methods (\sect\ref{sec:outcomes}), it is not perfect. The large volume of data of data used in our analyses minimizes the impact of any single misclassification by the model, and all of our figures include bootstrapped CIs to better understand the robustness of our findings. Furthermore, systems as complex as online communities have countless confounding factors that can bias analyses such as ours. While we attempt to control for confounders by using IPTW and DID causal inference methods (\sect\ref{sec:iptw_did}), future work should make use of active experimentation for gold-standard causal estimates.

Validating measures of online governance is challenging, as there is minimal `ground truth' to use for assessment. While it exceeds the scope of this work, a large-scale survey of many community members' perceptions of their moderators could be used to refine and validate future models. Future models could go beyond just positive and negative sentiment to identity biased, overly-strict, or too-permissive moderation, for example.

While our methods for quantifying communities' practices are sophisticated, they also miss many key aspects of governance. 
Our analyses of moderators' engagement with communities (\sect\ref{subsec:mod_engagement}) only consider how much each moderator posts in their communities, not \emph{what} they post. Future analyses could examine the types of contributions that moderators make. Our analyses of moderator team dynamics are also limited by the lack of publicly available data about which moderators are active; many communities' mod teams contain mods who do not contribute to the day-to-day governance of the community. Future work should incorporate detailed information about specific moderators' actions, although data collection is a substantial challenge \cite{li_2022_modlogs}.

\subsection{Conclusion}\label{sec:conclusion}
Good governance is critical to the functioning of online communities, yet it is difficult to know what governance practices are most effective, as it is challenging to measure the `success' of community governance. In this work, we developed a method to quantify community members' perceptions of their moderators across thousands of communities. We relate these perceptions to different aspects of governance including community size and topic (\sect\ref{sec:rq1}) as well as to different actions that moderators can take, including rule enforcement (\sect\ref{subsec:modteam_size_actions}), community engagement (\sect\ref{subsec:mod_engagement}), and moderator recruiting practices (\sect\ref{subsec:mod_recruiting}-\ref{subsec:mod_experience}). We empirically identify promising strategies for community moderators, including tailoring the strictness of rule enforcement to the community topic, and recruiting engaged mods. We make our models and anonymized datasets public to support future research.
\begin{acks}
This research was supported by the Office of Naval Research (\#N00014-21-1-2154), NSF grant IIS-1901386, NSF CAREER IIS-2142794, NSF grant CNS-2025022, and the Bill \& Melinda Gates Foundation (INV-004841). This work was completed on Hyak, UW’s high performance computing cluster. This resource was funded by the UW student technology fee.
\end{acks}
\section{Ethical Considerations}\label{sec:ethics}
We believe this work will have a positive broader impact by informing better moderation practices in online communities, as well as providing researchers with better tools to study community members' perceptions of moderators. As we only make use of public data, we believe our work has minimal risk to participants' privacy. As research has shown that some online community users are uncomfortable with their data being used for research, even when posted publicly \cite{Fiesler2018ParticipantPO}, we take further steps to reduce potential harms and misuse potential of our mod discourse dataset: we do not publish usernames or identifiable information, only predicted sentiment with regards to the moderators. We publish moderator timelines, including moderators' usernames, however these usernames are already publicly listed on communities' `about' pages. Upon publication of this dataset, we will provide affordances for users to have their data removed at their request.
We comply with relevant licenses for NLP models we use or modify. 
This study was approved by the University of Washington IRB under ID number STUDY00011457.
\bibliographystyle{ACM-Reference-Format}
\bibliography{bibliography}


\begin{thebibliography}{47}


\ifx \showCODEN    \undefined \def \showCODEN     #1{\unskip}     \fi
\ifx \showDOI      \undefined \def \showDOI       #1{#1}\fi
\ifx \showISBNx    \undefined \def \showISBNx     #1{\unskip}     \fi
\ifx \showISBNxiii \undefined \def \showISBNxiii  #1{\unskip}     \fi
\ifx \showISSN     \undefined \def \showISSN      #1{\unskip}     \fi
\ifx \showLCCN     \undefined \def \showLCCN      #1{\unskip}     \fi
\ifx \shownote     \undefined \def \shownote      #1{#1}          \fi
\ifx \showarticletitle \undefined \def \showarticletitle #1{#1}   \fi
\ifx \showURL      \undefined \def \showURL       {\relax}        \fi
\providecommand\bibfield[2]{#2}
\providecommand\bibinfo[2]{#2}
\providecommand\natexlab[1]{#1}
\providecommand\showeprint[2][]{arXiv:#2}

\bibitem[int({[n.\,d.]})]%
        {internet_archive}
 \bibinfo{year}{[n.\,d.]}\natexlab{}.
\newblock \bibinfo{title}{Internet Archive}.
\newblock
\newblock
\urldef\tempurl%
\url{https://archive.org/}
\showURL{%
\tempurl}


\bibitem[Almerekhi et~al\mbox{.}(2020)]%
        {almerekhi_2020_mod_harassment_modeling}
\bibfield{author}{\bibinfo{person}{Hind Almerekhi}, \bibinfo{person}{Haewoon Kwak}, {and} \bibinfo{person}{Bernard~Jim Jansen}.} \bibinfo{year}{2020}\natexlab{}.
\newblock \showarticletitle{Statistical Modeling of Harassment against Reddit Moderators}.
\newblock \bibinfo{journal}{\emph{WWW Companion}} (\bibinfo{year}{2020}).
\newblock
\urldef\tempurl%
\url{https://api.semanticscholar.org/CorpusID:218522526}
\showURL{%
\tempurl}


\bibitem[Althoff et~al\mbox{.}(2016)]%
        {althoff2016quantifying}
\bibfield{author}{\bibinfo{person}{Tim Althoff}, \bibinfo{person}{R Sosic}, \bibinfo{person}{JL Hicks}, \bibinfo{person}{AC King}, \bibinfo{person}{SL Delp}, {and} \bibinfo{person}{J Leskovec}.} \bibinfo{year}{2016}\natexlab{}.
\newblock \showarticletitle{Quantifying dose response relationships between physical activity and health using propensity scores}.
\newblock \bibinfo{journal}{\emph{NIPS ML4H}} (\bibinfo{year}{2016}).
\newblock


\bibitem[Austin and Stuart(2015)]%
        {Austin2015_IPTW_best_practice}
\bibfield{author}{\bibinfo{person}{Peter~C. Austin} {and} \bibinfo{person}{Elizabeth~A. Stuart}.} \bibinfo{year}{2015}\natexlab{}.
\newblock \showarticletitle{Moving towards best practice when using inverse probability of treatment weighting (IPTW) using the propensity score to estimate causal treatment effects in observational studies}.
\newblock \bibinfo{journal}{\emph{Statistics in Medicine}}  \bibinfo{volume}{34} (\bibinfo{year}{2015}), \bibinfo{pages}{3661 -- 3679}.
\newblock
\urldef\tempurl%
\url{https://api.semanticscholar.org/CorpusID:14478957}
\showURL{%
\tempurl}


\bibitem[Bao et~al\mbox{.}(2021)]%
        {Bao_2021_prosocial}
\bibfield{author}{\bibinfo{person}{Jiajun Bao}, \bibinfo{person}{J. Wu}, \bibinfo{person}{Yiming Zhang}, \bibinfo{person}{Eshwar Chandrasekharan}, {and} \bibinfo{person}{David Jurgens}.} \bibinfo{year}{2021}\natexlab{}.
\newblock \showarticletitle{Conversations Gone Alright: Quantifying and Predicting Prosocial Outcomes in Online Conversations}.
\newblock \bibinfo{journal}{\emph{TheWebConf}} (\bibinfo{year}{2021}).
\newblock
\urldef\tempurl%
\url{https://api.semanticscholar.org/CorpusID:231933671}
\showURL{%
\tempurl}


\bibitem[Baumgartner et~al\mbox{.}(2020)]%
        {baumgartner_pushshift_2020}
\bibfield{author}{\bibinfo{person}{Jason Baumgartner}, \bibinfo{person}{Savvas Zannettou}, \bibinfo{person}{Brian Keegan}, \bibinfo{person}{Megan Squire}, {and} \bibinfo{person}{Jeremy Blackburn}.} \bibinfo{year}{2020}\natexlab{}.
\newblock \showarticletitle{The {Pushshift} {Reddit} {Dataset}}.
\newblock \bibinfo{journal}{\emph{arXiv:2001.08435}} (\bibinfo{year}{2020}).
\newblock


\bibitem[Callaway and Sant’Anna(2021)]%
        {callaway_2021_diff_in_diff}
\bibfield{author}{\bibinfo{person}{Brantly Callaway} {and} \bibinfo{person}{Pedro~H.C. Sant’Anna}.} \bibinfo{year}{2021}\natexlab{}.
\newblock \showarticletitle{Difference-in-Differences with multiple time periods}.
\newblock \bibinfo{journal}{\emph{Journal of Econometrics}} \bibinfo{volume}{225}, \bibinfo{number}{2} (\bibinfo{year}{2021}), \bibinfo{pages}{200--230}.
\newblock


\bibitem[Chandrasekharan et~al\mbox{.}(2020)]%
        {Chandrasekharan2020Quarantined}
\bibfield{author}{\bibinfo{person}{Eshwar Chandrasekharan}, \bibinfo{person}{Shagun Jhaver}, \bibinfo{person}{Amy Bruckman}, {and} \bibinfo{person}{Eric Gilbert}.} \bibinfo{year}{2020}\natexlab{}.
\newblock \showarticletitle{Quarantined! Examining the Effects of a Community-Wide Moderation Intervention on Reddit}.
\newblock \bibinfo{journal}{\emph{TOCHI}}  \bibinfo{volume}{29} (\bibinfo{year}{2020}), \bibinfo{pages}{1 -- 26}.
\newblock
\urldef\tempurl%
\url{https://api.semanticscholar.org/CorpusID:221879027}
\showURL{%
\tempurl}


\bibitem[Chandrasekharan et~al\mbox{.}(2018)]%
        {chandrasekharan_internets_2018}
\bibfield{author}{\bibinfo{person}{Eshwar Chandrasekharan}, \bibinfo{person}{Mattia Samory}, \bibinfo{person}{Shagun Jhaver}, \bibinfo{person}{Hunter Charvat}, \bibinfo{person}{Amy Bruckman}, \bibinfo{person}{Cliff Lampe}, \bibinfo{person}{Jacob Eisenstein}, {and} \bibinfo{person}{Eric Gilbert}.} \bibinfo{year}{2018}\natexlab{}.
\newblock \showarticletitle{The {Internet}'s {Hidden} {Rules}: {An} {Empirical} {Study} of {Reddit} {Norm} {Violations} at {Micro}, {Meso}, and {Macro} {Scales}}.
\newblock \bibinfo{journal}{\emph{CSCW}} (\bibinfo{year}{2018}).
\newblock


\bibitem[Danescu-Niculescu-Mizil et~al\mbox{.}(2013)]%
        {dnm_2013_no_country}
\bibfield{author}{\bibinfo{person}{Cristian Danescu-Niculescu-Mizil}, \bibinfo{person}{Robert West}, \bibinfo{person}{Dan Jurafsky}, \bibinfo{person}{Jure Leskovec}, {and} \bibinfo{person}{Christopher Potts}.} \bibinfo{year}{2013}\natexlab{}.
\newblock \showarticletitle{No country for old members: user lifecycle and linguistic change in online communities}.
\newblock \bibinfo{journal}{\emph{WWW}} (\bibinfo{year}{2013}).
\newblock
\urldef\tempurl%
\url{https://api.semanticscholar.org/CorpusID:14466332}
\showURL{%
\tempurl}


\bibitem[Dettmers et~al\mbox{.}(2023)]%
        {dettmers2023QLoRA}
\bibfield{author}{\bibinfo{person}{Tim Dettmers}, \bibinfo{person}{Artidoro Pagnoni}, \bibinfo{person}{Ari Holtzman}, {and} \bibinfo{person}{Luke Zettlemoyer}.} \bibinfo{year}{2023}\natexlab{}.
\newblock \showarticletitle{QLoRA: Efficient Finetuning of Quantized LLMs}.
\newblock \bibinfo{journal}{\emph{ArXiv}}  \bibinfo{volume}{abs/2305.14314} (\bibinfo{year}{2023}).
\newblock
\urldef\tempurl%
\url{https://api.semanticscholar.org/CorpusID:258841328}
\showURL{%
\tempurl}


\bibitem[Fiesler et~al\mbox{.}(2018)]%
        {fiesler_2018_reddit_rules}
\bibfield{author}{\bibinfo{person}{Casey Fiesler}, \bibinfo{person}{Jialun~Aaron Jiang}, \bibinfo{person}{Joshua McCann}, \bibinfo{person}{Kyle Frye}, {and} \bibinfo{person}{Jed~R. Brubaker}.} \bibinfo{year}{2018}\natexlab{}.
\newblock \showarticletitle{Reddit Rules! Characterizing an Ecosystem of Governance}.
\newblock \bibinfo{journal}{\emph{ICWSM}} (\bibinfo{year}{2018}).
\newblock
\urldef\tempurl%
\url{https://api.semanticscholar.org/CorpusID:49404607}
\showURL{%
\tempurl}


\bibitem[Fiesler and Proferes(2018)]%
        {Fiesler2018ParticipantPO}
\bibfield{author}{\bibinfo{person}{Casey Fiesler} {and} \bibinfo{person}{Nicholas Proferes}.} \bibinfo{year}{2018}\natexlab{}.
\newblock \showarticletitle{“Participant” Perceptions of Twitter Research Ethics}.
\newblock \bibinfo{journal}{\emph{Social Media + Society}}  \bibinfo{volume}{4} (\bibinfo{year}{2018}).
\newblock


\bibitem[Frey and Sumner(2019)]%
        {frey_2019_minecraft_gov}
\bibfield{author}{\bibinfo{person}{Seth Frey} {and} \bibinfo{person}{Robert~W. Sumner}.} \bibinfo{year}{2019}\natexlab{}.
\newblock \showarticletitle{Emergence of integrated institutions in a large population of self-governing communities}.
\newblock \bibinfo{journal}{\emph{PLOS ONE}} \bibinfo{volume}{14}, \bibinfo{number}{7} (\bibinfo{date}{07} \bibinfo{year}{2019}), \bibinfo{pages}{1--18}.
\newblock
\urldef\tempurl%
\url{https://doi.org/10.1371/journal.pone.0216335}
\showDOI{\tempurl}


\bibitem[Hutto and Gilbert(2014)]%
        {hutto_2014_VADER}
\bibfield{author}{\bibinfo{person}{Clayton~J. Hutto} {and} \bibinfo{person}{Eric Gilbert}.} \bibinfo{year}{2014}\natexlab{}.
\newblock \showarticletitle{VADER: A Parsimonious Rule-Based Model for Sentiment Analysis of Social Media Text}.
\newblock \bibinfo{journal}{\emph{ICWSM}} (\bibinfo{year}{2014}).
\newblock
\urldef\tempurl%
\url{https://api.semanticscholar.org/CorpusID:12233345}
\showURL{%
\tempurl}


\bibitem[Hwang and Foote(2021)]%
        {hwang_2021_small_communities}
\bibfield{author}{\bibinfo{person}{Sohyeon Hwang} {and} \bibinfo{person}{Jeremy~D. Foote}.} \bibinfo{year}{2021}\natexlab{}.
\newblock \showarticletitle{Why Do People Participate in Small Online Communities?}
\newblock \bibinfo{journal}{\emph{CSCW}} (\bibinfo{year}{2021}).
\newblock
\urldef\tempurl%
\url{https://doi.org/10.1145/3479606}
\showURL{%
\tempurl}


\bibitem[Hwang and Shaw(2022)]%
        {Hwang2022_wikipedia_rules}
\bibfield{author}{\bibinfo{person}{Sohyeon Hwang} {and} \bibinfo{person}{Aaron Shaw}.} \bibinfo{year}{2022}\natexlab{}.
\newblock \showarticletitle{Rules and Rule-Making in the Five Largest Wikipedias}. In \bibinfo{booktitle}{\emph{ICWSM}}.
\newblock
\urldef\tempurl%
\url{https://api.semanticscholar.org/CorpusID:249892728}
\showURL{%
\tempurl}


\bibitem[Jhaver et~al\mbox{.}(2019)]%
        {jhaver_2019_removal_reasons}
\bibfield{author}{\bibinfo{person}{Shagun Jhaver}, \bibinfo{person}{Amy Bruckman}, {and} \bibinfo{person}{Eric Gilbert}.} \bibinfo{year}{2019}\natexlab{}.
\newblock \showarticletitle{Does Transparency in Moderation Really Matter?}
\newblock \bibinfo{journal}{\emph{CSCW}}  \bibinfo{volume}{3} (\bibinfo{year}{2019}), \bibinfo{pages}{1 -- 27}.
\newblock
\urldef\tempurl%
\url{https://api.semanticscholar.org/CorpusID:203558216}
\showURL{%
\tempurl}


\bibitem[Jhaver et~al\mbox{.}(2023a)]%
        {jhaver2023decentralizing}
\bibfield{author}{\bibinfo{person}{Shagun Jhaver}, \bibinfo{person}{Seth Frey}, {and} \bibinfo{person}{Amy Zhang}.} \bibinfo{year}{2023}\natexlab{a}.
\newblock \bibinfo{title}{Decentralizing Platform Power: A Design Space of Multi-level Governance in Online Social Platforms}.
\newblock
\newblock
\showeprint[arxiv]{2108.12529}~[cs.HC]


\bibitem[Jhaver et~al\mbox{.}(2023b)]%
        {jhaver2023bystanders}
\bibfield{author}{\bibinfo{person}{Shagun Jhaver}, \bibinfo{person}{Himanshu Rathi}, {and} \bibinfo{person}{Koustuv Saha}.} \bibinfo{year}{2023}\natexlab{b}.
\newblock \bibinfo{title}{Bystanders of Online Moderation: Examining the Effects of Witnessing Post-Removal Explanations}.
\newblock
\newblock
\showeprint[arxiv]{2309.08361}~[cs.HC]


\bibitem[Kheiri and Karimi(2023)]%
        {kheiri_2023_SentimentGPT}
\bibfield{author}{\bibinfo{person}{Kiana Kheiri} {and} \bibinfo{person}{Hamid Karimi}.} \bibinfo{year}{2023}\natexlab{}.
\newblock \showarticletitle{SentimentGPT: Exploiting GPT for Advanced Sentiment Analysis and its Departure from Current Machine Learning}.
\newblock \bibinfo{journal}{\emph{ArXiv}}  \bibinfo{volume}{abs/2307.10234} (\bibinfo{year}{2023}).
\newblock
\urldef\tempurl%
\url{https://api.semanticscholar.org/CorpusID:259991148}
\showURL{%
\tempurl}


\bibitem[Koshy et~al\mbox{.}(2023)]%
        {koshy_2023_user_mod_alignment}
\bibfield{author}{\bibinfo{person}{Vinay Koshy}, \bibinfo{person}{Tanvi Bajpai}, \bibinfo{person}{Eshwar Chandrasekharan}, \bibinfo{person}{Hari Sundaram}, {and} \bibinfo{person}{Karrie Karahalios}.} \bibinfo{year}{2023}\natexlab{}.
\newblock \showarticletitle{Measuring User-Moderator Alignment on r/ChangeMyView}.
\newblock \bibinfo{journal}{\emph{CSCW}}  \bibinfo{volume}{7} (\bibinfo{year}{2023}), \bibinfo{pages}{1 -- 36}.
\newblock
\urldef\tempurl%
\url{https://api.semanticscholar.org/CorpusID:263621128}
\showURL{%
\tempurl}


\bibitem[Kraut and Resnick(2012)]%
        {kraut2012building}
\bibfield{author}{\bibinfo{person}{Robert~E Kraut} {and} \bibinfo{person}{Paul Resnick}.} \bibinfo{year}{2012}\natexlab{}.
\newblock \bibinfo{booktitle}{\emph{Building successful online communities: Evidence-based social design}}.
\newblock \bibinfo{publisher}{Mit Press}.
\newblock


\bibitem[Kumar et~al\mbox{.}(2023)]%
        {Kumar2023WatchYL}
\bibfield{author}{\bibinfo{person}{Deepak Kumar}, \bibinfo{person}{Yousef~Anees AbuHashem}, {and} \bibinfo{person}{Zakir Durumeric}.} \bibinfo{year}{2023}\natexlab{}.
\newblock \showarticletitle{Watch Your Language: Large Language Models and Content Moderation}. In \bibinfo{booktitle}{\emph{International Conference on Web and Social Media}}.
\newblock
\urldef\tempurl%
\url{https://api.semanticscholar.org/CorpusID:262825535}
\showURL{%
\tempurl}


\bibitem[Landis and Koch(1977)]%
        {landis_koch_1977_measurement}
\bibfield{author}{\bibinfo{person}{J.~Richard Landis} {and} \bibinfo{person}{Gary~G. Koch}.} \bibinfo{year}{1977}\natexlab{}.
\newblock \showarticletitle{The Measurement of Observer Agreement for Categorical Data}.
\newblock \bibinfo{journal}{\emph{Biometrics}} (\bibinfo{year}{1977}).
\newblock


\bibitem[Li et~al\mbox{.}(2022a)]%
        {li_2022_modlogs}
\bibfield{author}{\bibinfo{person}{Hanlin Li}, \bibinfo{person}{Brent~J. Hecht}, {and} \bibinfo{person}{Stevie Chancellor}.} \bibinfo{year}{2022}\natexlab{a}.
\newblock \showarticletitle{All That's Happening behind the Scenes: Putting the Spotlight on Volunteer Moderator Labor in Reddit}.
\newblock \bibinfo{journal}{\emph{ICWSM}} (\bibinfo{year}{2022}).
\newblock
\urldef\tempurl%
\url{https://api.semanticscholar.org/CorpusID:249191685}
\showURL{%
\tempurl}


\bibitem[Li et~al\mbox{.}(2022b)]%
        {li_measuring_2022}
\bibfield{author}{\bibinfo{person}{Hanlin Li}, \bibinfo{person}{Brent~J. Hecht}, {and} \bibinfo{person}{Stevie Chancellor}.} \bibinfo{year}{2022}\natexlab{b}.
\newblock \showarticletitle{Measuring the Monetary Value of Online Volunteer Work}.
\newblock \bibinfo{journal}{\emph{ICWSM}} (\bibinfo{year}{2022}).
\newblock


\bibitem[Lin et~al\mbox{.}(2017)]%
        {Lin_2017_Better_smaller}
\bibfield{author}{\bibinfo{person}{Zhiyuan~Jerry Lin}, \bibinfo{person}{Niloufar Salehi}, \bibinfo{person}{Bowen Yao}, \bibinfo{person}{Yiqi Chen}, {and} \bibinfo{person}{Michael~S. Bernstein}.} \bibinfo{year}{2017}\natexlab{}.
\newblock \showarticletitle{Better When It Was Smaller? Community Content and Behavior After Massive Growth}. In \bibinfo{booktitle}{\emph{ICWSM}}.
\newblock
\urldef\tempurl%
\url{https://api.semanticscholar.org/CorpusID:2524208}
\showURL{%
\tempurl}


\bibitem[Liu et~al\mbox{.}(2019)]%
        {Liu_2019_RoBERTa}
\bibfield{author}{\bibinfo{person}{Yinhan Liu}, \bibinfo{person}{Myle Ott}, \bibinfo{person}{Naman Goyal}, \bibinfo{person}{Jingfei Du}, \bibinfo{person}{Mandar Joshi}, \bibinfo{person}{Danqi Chen}, \bibinfo{person}{Omer Levy}, \bibinfo{person}{Mike Lewis}, \bibinfo{person}{Luke Zettlemoyer}, {and} \bibinfo{person}{Veselin Stoyanov}.} \bibinfo{year}{2019}\natexlab{}.
\newblock \showarticletitle{RoBERTa: A Robustly Optimized BERT Pretraining Approach}.
\newblock \bibinfo{journal}{\emph{ArXiv}}  \bibinfo{volume}{abs/1907.11692} (\bibinfo{year}{2019}).
\newblock
\urldef\tempurl%
\url{https://api.semanticscholar.org/CorpusID:198953378}
\showURL{%
\tempurl}


\bibitem[Matias(2019a)]%
        {matias_civic_2019}
\bibfield{author}{\bibinfo{person}{J.~Nathan Matias}.} \bibinfo{year}{2019}\natexlab{a}.
\newblock \showarticletitle{The {Civic} {Labor} of {Volunteer} {Moderators} {Online}}.
\newblock \bibinfo{journal}{\emph{Social Media + Society}} (\bibinfo{year}{2019}).
\newblock


\bibitem[Matias(2019b)]%
        {Matias_2019_Preventing}
\bibfield{author}{\bibinfo{person}{Jorge~Nathan Matias}.} \bibinfo{year}{2019}\natexlab{b}.
\newblock \showarticletitle{Preventing harassment and increasing group participation through social norms in 2,190 online science discussions}.
\newblock \bibinfo{journal}{\emph{PNAS}}  \bibinfo{volume}{116} (\bibinfo{year}{2019}), \bibinfo{pages}{9785 -- 9789}.
\newblock
\urldef\tempurl%
\url{https://api.semanticscholar.org/CorpusID:140369949}
\showURL{%
\tempurl}


\bibitem[OpenAI et~al\mbox{.}(2023)]%
        {achiam_2023_GPT4}
\bibfield{author}{\bibinfo{person}{OpenAI} {et~al\mbox{.}}} \bibinfo{year}{2023}\natexlab{}.
\newblock \bibinfo{title}{GPT-4 Technical Report}.
\newblock
\newblock
\showeprint[arxiv]{2303.08774}~[cs.CL]


\bibitem[Panciera et~al\mbox{.}(2009)]%
        {Panciera2009WikipediansAB}
\bibfield{author}{\bibinfo{person}{Katherine~A. Panciera}, \bibinfo{person}{Aaron~L Halfaker}, {and} \bibinfo{person}{Loren~G. Terveen}.} \bibinfo{year}{2009}\natexlab{}.
\newblock \showarticletitle{Wikipedians are born, not made: a study of power editors on Wikipedia}.
\newblock \bibinfo{journal}{\emph{GROUP}} (\bibinfo{year}{2009}).
\newblock
\urldef\tempurl%
\url{https://api.semanticscholar.org/CorpusID:6286454}
\showURL{%
\tempurl}


\bibitem[Proferes et~al\mbox{.}(2021)]%
        {Proferes_2021_reddit_research_overview}
\bibfield{author}{\bibinfo{person}{Nicholas Proferes}, \bibinfo{person}{Naiyan Jones}, \bibinfo{person}{Sarah~A. Gilbert}, \bibinfo{person}{Casey Fiesler}, {and} \bibinfo{person}{Michael Zimmer}.} \bibinfo{year}{2021}\natexlab{}.
\newblock \showarticletitle{Studying Reddit: A Systematic Overview of Disciplines, Approaches, Methods, and Ethics}.
\newblock \bibinfo{journal}{\emph{Social Media + Society}}  \bibinfo{volume}{7} (\bibinfo{year}{2021}).
\newblock
\urldef\tempurl%
\url{https://api.semanticscholar.org/CorpusID:235724916}
\showURL{%
\tempurl}


\bibitem[Reddy and Chandrasekharan(2023)]%
        {reddy_2023_evolution_rules}
\bibfield{author}{\bibinfo{person}{Harita Reddy} {and} \bibinfo{person}{Eshwar Chandrasekharan}.} \bibinfo{year}{2023}\natexlab{}.
\newblock \showarticletitle{Evolution of Rules in Reddit Communities}.
\newblock \bibinfo{journal}{\emph{CSCW}} (\bibinfo{year}{2023}).
\newblock
\urldef\tempurl%
\url{https://api.semanticscholar.org/CorpusID:264039135}
\showURL{%
\tempurl}


\bibitem[Ribeiro et~al\mbox{.}(2020)]%
        {Ribeiro2020_migration}
\bibfield{author}{\bibinfo{person}{Manoel~Horta Ribeiro}, \bibinfo{person}{Shagun Jhaver}, \bibinfo{person}{Savvas Zannettou}, \bibinfo{person}{Jeremy Blackburn}, \bibinfo{person}{Emiliano~De Cristofaro}, \bibinfo{person}{Gianluca Stringhini}, {and} \bibinfo{person}{Robert West}.} \bibinfo{year}{2020}\natexlab{}.
\newblock \showarticletitle{Do Platform Migrations Compromise Content Moderation? Evidence from r/The\_Donald and r/Incels}.
\newblock \bibinfo{journal}{\emph{CSCW}} (\bibinfo{year}{2020}).
\newblock
\urldef\tempurl%
\url{https://api.semanticscholar.org/CorpusID:263873469}
\showURL{%
\tempurl}


\bibitem[Seering and Kairam(2022)]%
        {seering_2022_twitch_moderator_recruiting}
\bibfield{author}{\bibinfo{person}{Joseph Seering} {and} \bibinfo{person}{Sanjay~Ram Kairam}.} \bibinfo{year}{2022}\natexlab{}.
\newblock \showarticletitle{Who Moderates on Twitch and What Do They Do?}
\newblock \bibinfo{journal}{\emph{GROUP}}  \bibinfo{volume}{7} (\bibinfo{year}{2022}), \bibinfo{pages}{1 -- 18}.
\newblock
\urldef\tempurl%
\url{https://api.semanticscholar.org/CorpusID:255226737}
\showURL{%
\tempurl}


\bibitem[Seering et~al\mbox{.}(2019)]%
        {Seering2019ModeratorEA}
\bibfield{author}{\bibinfo{person}{Joseph Seering}, \bibinfo{person}{Tonya Wang}, \bibinfo{person}{Jina Yoon}, {and} \bibinfo{person}{Geoff~F. Kaufman}.} \bibinfo{year}{2019}\natexlab{}.
\newblock \showarticletitle{Moderator engagement and community development in the age of algorithms}.
\newblock \bibinfo{journal}{\emph{New Media \& Society}}  \bibinfo{volume}{21} (\bibinfo{year}{2019}), \bibinfo{pages}{1417 -- 1443}.
\newblock
\urldef\tempurl%
\url{https://api.semanticscholar.org/CorpusID:149757447}
\showURL{%
\tempurl}


\bibitem[Srinivasan et~al\mbox{.}(2019)]%
        {srinivasan_2019_content_removal_cmv}
\bibfield{author}{\bibinfo{person}{Kumar~Bhargav Srinivasan}, \bibinfo{person}{Cristian Danescu-Niculescu-Mizil}, \bibinfo{person}{Lillian Lee}, {and} \bibinfo{person}{Chenhao Tan}.} \bibinfo{year}{2019}\natexlab{}.
\newblock \showarticletitle{Content Removal as a Moderation Strategy}.
\newblock \bibinfo{journal}{\emph{CSCW}} (\bibinfo{year}{2019}).
\newblock
\urldef\tempurl%
\url{https://api.semanticscholar.org/CorpusID:203617040}
\showURL{%
\tempurl}


\bibitem[Thomas et~al\mbox{.}(2021)]%
        {Thomas_2021_bans_behavior}
\bibfield{author}{\bibinfo{person}{Pamela~Bilo Thomas}, \bibinfo{person}{Daniel Riehm}, \bibinfo{person}{Maria Glenski}, {and} \bibinfo{person}{Tim Weninger}.} \bibinfo{year}{2021}\natexlab{}.
\newblock \showarticletitle{Behavior Change in Response to Subreddit Bans and External Events}.
\newblock \bibinfo{journal}{\emph{IEEE TCSS}}  \bibinfo{volume}{8} (\bibinfo{year}{2021}), \bibinfo{pages}{809--818}.
\newblock
\urldef\tempurl%
\url{https://api.semanticscholar.org/CorpusID:230770317}
\showURL{%
\tempurl}


\bibitem[Touvron et~al\mbox{.}(2023)]%
        {touvron2023llama2}
\bibfield{author}{\bibinfo{person}{Hugo Touvron} {et~al\mbox{.}}} \bibinfo{year}{2023}\natexlab{}.
\newblock \bibinfo{title}{Llama 2: Open Foundation and Fine-Tuned Chat Models}.
\newblock
\newblock
\showeprint[arxiv]{2307.09288}~[cs.CL]


\bibitem[Waller and Anderson(2020)]%
        {Waller_2020_embeddings}
\bibfield{author}{\bibinfo{person}{Isaac Waller} {and} \bibinfo{person}{Ashton Anderson}.} \bibinfo{year}{2020}\natexlab{}.
\newblock \showarticletitle{Quantifying social organization and political polarization in online platforms}.
\newblock \bibinfo{journal}{\emph{Nature}} (\bibinfo{year}{2020}).
\newblock
\urldef\tempurl%
\url{https://api.semanticscholar.org/CorpusID:236469369}
\showURL{%
\tempurl}


\bibitem[Weld et~al\mbox{.}(2021)]%
        {weld_2021_news_sharing}
\bibfield{author}{\bibinfo{person}{Galen~Cassebeer Weld}, \bibinfo{person}{Maria Glenski}, {and} \bibinfo{person}{Tim Althoff}.} \bibinfo{year}{2021}\natexlab{}.
\newblock \showarticletitle{Political Bias and Factualness in News Sharing Across more then 100, 000 Online Communities}.
\newblock \bibinfo{journal}{\emph{ICWSM}} (\bibinfo{year}{2021}).
\newblock
\urldef\tempurl%
\url{https://api.semanticscholar.org/CorpusID:231942492}
\showURL{%
\tempurl}


\bibitem[Weld et~al\mbox{.}(2022)]%
        {weld_2022_survey_icwsm}
\bibfield{author}{\bibinfo{person}{Galen~Cassebeer Weld}, \bibinfo{person}{Amy~X. Zhang}, {and} \bibinfo{person}{Tim Althoff}.} \bibinfo{year}{2022}\natexlab{}.
\newblock \showarticletitle{What Makes Online Communities 'Better'? Measuring Values, Consensus, and Conflict across Thousands of Subreddits}.
\newblock \bibinfo{journal}{\emph{ICWSM}} (\bibinfo{year}{2022}).
\newblock


\bibitem[Weld et~al\mbox{.}(2024)]%
        {weld_values_taxonomy_2021}
\bibfield{author}{\bibinfo{person}{Galen~Cassebeer Weld}, \bibinfo{person}{Amy~X. Zhang}, {and} \bibinfo{person}{Tim Althoff}.} \bibinfo{year}{2024}\natexlab{}.
\newblock \showarticletitle{Making Online Communities 'Better': A Taxonomy of Community Values on Reddit}.
\newblock \bibinfo{journal}{\emph{ICWSM}} (\bibinfo{year}{2024}).
\newblock


\bibitem[Zhang et~al\mbox{.}(2020)]%
        {zhang_policykit_2020}
\bibfield{author}{\bibinfo{person}{Amy~X. Zhang}, \bibinfo{person}{Grant Hugh}, {and} \bibinfo{person}{Michael~S. Bernstein}.} \bibinfo{year}{2020}\natexlab{}.
\newblock \showarticletitle{{PolicyKit}: {Building} {Governance} in {Online} {Communities}}.
\newblock \bibinfo{journal}{\emph{UIST}} (\bibinfo{year}{2020}).
\newblock


\bibitem[Zhu et~al\mbox{.}(2014)]%
        {zhu_2014_wikia_membership}
\bibfield{author}{\bibinfo{person}{Haiyi Zhu}, \bibinfo{person}{Robert~E. Kraut}, {and} \bibinfo{person}{Aniket Kittur}.} \bibinfo{year}{2014}\natexlab{}.
\newblock \showarticletitle{The Impact of Membership Overlap on the Survival of Online Communities}.
\newblock \bibinfo{journal}{\emph{CHI}} (\bibinfo{year}{2014}), \bibinfo{pages}{281–290}.
\newblock
\urldef\tempurl%
\url{https://doi.org/10.1145/2556288.2557213}
\showURL{%
\tempurl}


\end{thebibliography}

\clearpage
\onecolumn
\appendix

\section{Moderator Sentiment Codebook}\label{app:codebook}
\subsection{Positive Sentiment}
This label should be used for comments expressing a positive sentiment towards the moderator or moderator team.

\xhdr{Examples}
\noindent
``This subreddit is so lucky to have such a great mod team''

\noindent
``Make the life of our hard-working mods here easier''

\noindent
``The mods are always so helpful, but this thread got a bit messy'' --- this is a tricky judgment call, but I’d say that the overall sentiment is positive with this thread being an exception.

\xhdr{Counterexample}
\noindent
``This subreddit used to be well-run, but in the past year or so the moderation has really gone to shit'' --- a judgment call similar to earlier, but here I would say \textit{negative}.

\subsection{Negative Sentiment}
This label should be used for comments expressing a negative sentiment towards the moderator or moderator team.

\xhdr{Examples}
\noindent
``The mods here suck''

\noindent
``The mods made a mistake'' --- everyone makes mistakes, but it’s still better if they don't.

\noindent
``I’m so tired of mods not removing crap like this''

\subsection{Neutral Sentiment}
This label should be used when there isn’t enough context for you to make a judgment about the sentiment of the comment or post, or the sentiment seems neutral.

\xhdr{Examples}
\noindent
``I didn't delete the post, maybe the mods did?''

\noindent
``Edit: reworded a slur after getting a warning from the mods''

\noindent
``Mods please ban this person'' --- not enough explicitly stated sentiment to know what is meant with certainty.

\noindent
``Why don’t you go and complain to the mods like you usually do?'' --- negative sentiment, but not directed towards the moderators.

\newpage
\section{Prompts for Topic and Sentiment Classification}\label{app:prompts}

\subsection{}\label{app:topic_prompt}
The following few-shot prompt was used with GPT-4~\cite{achiam_2023_GPT4} to classify the topic of subreddits in our analyses (\sect\ref{sec:topic_classifier}), based on the name of the subreddit.

\begin{lstlisting}

Given a subreddit, classify its topic into exactly one
of the following categories:

Hobby communities, which focus on a specific hobby,
sports related topic, or pastime.

Discussion communities, which are for discussion and
text-based content.

Media communities, which are for sharing videos,
and pictures.

News communities, which are for sharing news and
similar content.

Meme communites, which are for sharing memes or low
effort content.

Identity communities, which are for groups of a
spcecific identity or background.

/r/india: Identity
/r/bicycling,: Hobby
/r/CrappyDesign: Media
/r/me_irl: Memes
/r/worldnews: News
/r/AskReddit: Discussion
/r/nba: Hobby
/r/relationship_advice: Discussion
/r/science: News
/r/teenagers: Identity
/r/dankmemes: Memes
/r/pics: Media
/r/{subreddit}:

\end{lstlisting}

\newpage
\subsection{Sentiment Classification Step Prompt}\label{app:sentiment_prompt}
To classify the sentiment with regards to the moderators of posts and comments discussing mods (\sect\ref{sec:outcomes}), we used the following prompt for our LLaMA 2 model, fine tuned with QLoRA \cite{touvron2023llama2, dettmers2023QLoRA}.

\begin{lstlisting}
<|im_start|>system

Given a comment from Reddit which discusses moderators (or mods),
and its parent, identify how the author of the comment feels about
the moderators of the subreddit the comment was made in.

If the author feels that the moderators are doing a good job, mark
the sentiment as positive. If the author feels the moderators are
doing a bad job, mark the sentiment as negative. If it's not
possible to tell, mark the sentiment as neutral.


If the comment does not discuss moderators, but instead discusses
video game mods, or other types of modifications, mark the exclude
field as true and the sentiment as undefined.

If the comment is discussing moderators, but in a different
subreddit or different community, mark the other community
field as true.

The parent might be able to help your identification by providing
additional context.

Your answer should follow the format given in the examples.
<|im_end|>

<|im_start|>user
/r/{subreddit}
parent : {parent}
comment : {body}
<|im_end|>
\end{lstlisting}

\newpage
\section{Supplementary Figures}\label{app:additional_figs}

\begin{figure*}[h]
    \centering
    \includegraphics[width=\textwidth]{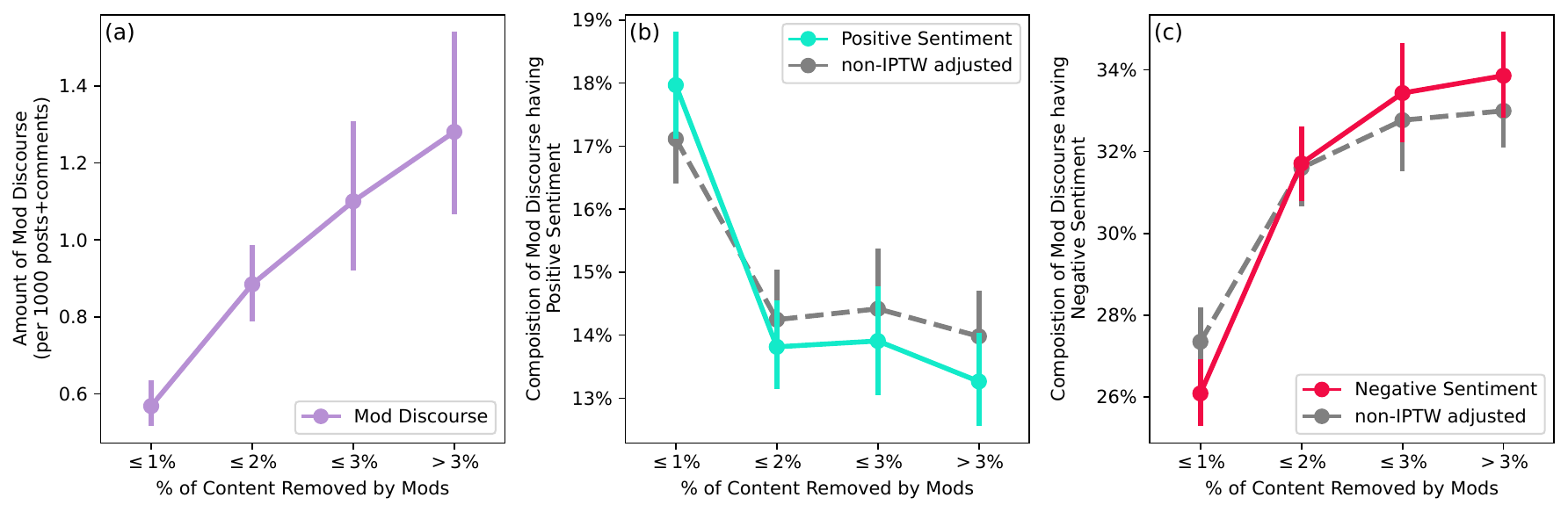}
    \caption{On average, communities with a larger fraction of their content removed by mods tend to have more a smaller fraction of their mod discourse have positive sentiment (b), and a larger fraction with negative sentiment (c). Communities with more content removed by mods also tend to have more total mod discourse (a).}
    \label{fig:removed_content}
\end{figure*}

\begin{figure*}[b]
    \centering
    \includegraphics[width=\textwidth]{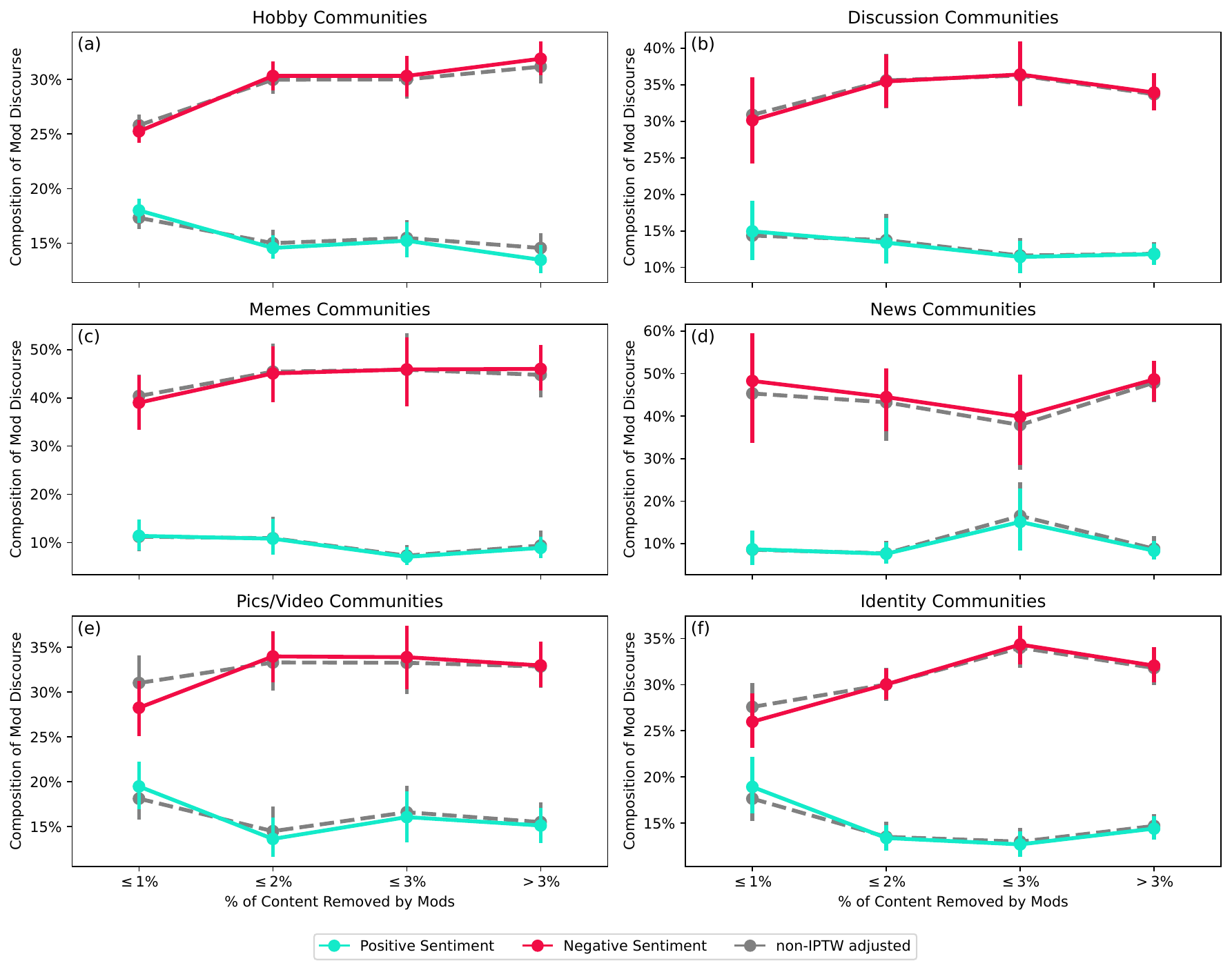}
    \caption{Generally, the fraction of mod discourse with negative sentiment is higher in communities with more removed content. However, these trends vary depending on the topic of the community.}
    \label{fig:removed_content_stratified_all_topics}
\end{figure*}

\begin{figure*}[b]
    \centering
    \includegraphics[width=.7\textwidth]{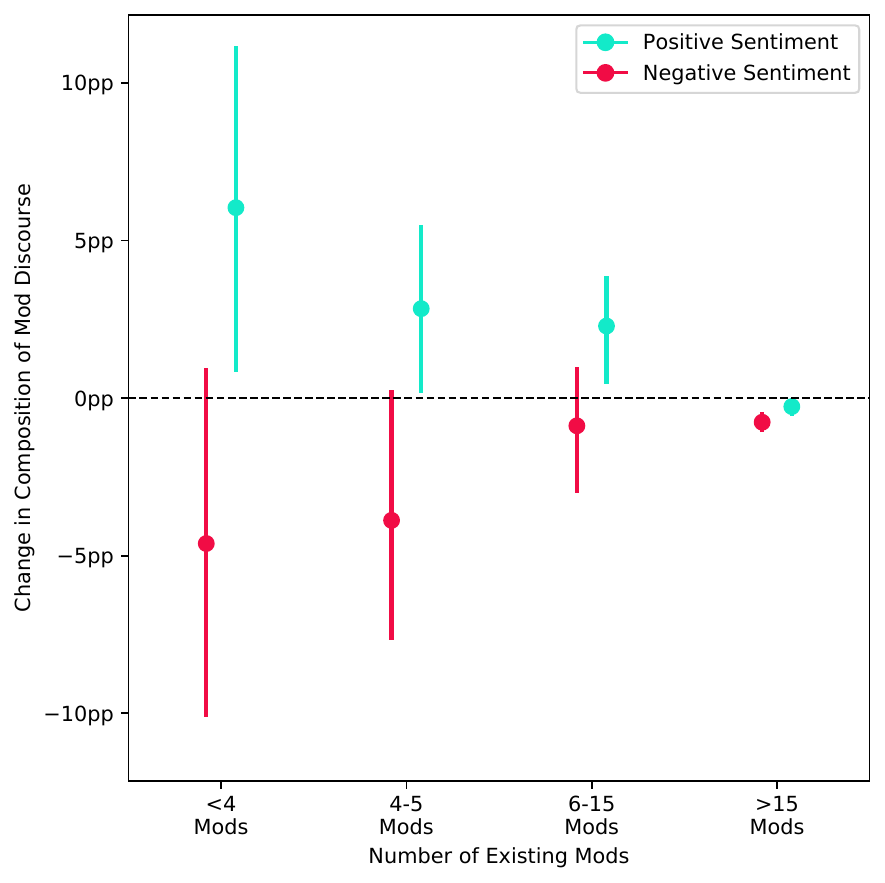}
    \caption{On average, adding a new moderator to a subreddit results in an increase in the fraction of mod discourse which has positive sentiment, and a corresponding decrease in the fraction that has negative sentiment. However, the magnitude of the impact varies with the size of the community's moderator team; adding a single new mod to a community with fewer than 4 mods has a much larger impact than adding a single mod to a community with more mods. Adding a single moderator to a community with more than 15 mods has an impact which is not significantly different from 0.}
    \label{fig:num_mods_appointment}
\end{figure*}

\clearpage
\section{Covariates Used in Propensity Score Modeling for IPTW}\label{app:covariates}

\subsection{Covariates used for Mod Workload Analyses}\label{app:covariates_workload}
The following table shows the covariates that were used in a logistic regression propensity score model for the moderator workload analyses (\fig\ref{fig:num_mods}), along with the resulting Standardized Mean Differences (SMDs) after reweighting, and the associated propensity score model (P.S.) coefficients used in the logistic regression.

{\scriptsize
\begin{tabular}{l|rr|rr|rr|rr|}
                               & \multicolumn{8}{c|}{Moderator Workload Treatment Bin (Posts+Comments per Mod per Day)}                                             \\
\multicolumn{1}{l|}{}          & \multicolumn{2}{c|}{0-5}     & \multicolumn{2}{c|}{5-10}    & \multicolumn{2}{c|}{10-100}  & \multicolumn{2}{c|}{>100}             \\ \hline
                               & SMD       & P.S. Coefficient & SMD       & P.S. Coefficient & SMD       & P.S. Coefficient & SMD            & P.S. Coefficient     \\
Covariate                      &           &                  &           &                  &           &                  &                &                      \\ \hline
\texttt{total\_items}          & -0.18     & -0.55            & -0.17     & -0.45            & -0.10     & +0.07            & +0.48          & +0.93                \\
\texttt{frac\_deleted}         & +0.05     & +0.01            & +0.02     & +0.02            & -0.07     & -0.07            & +0.20          & +0.04                \\
\texttt{frac\_removed}         & +0.19     & +0.04            & +0.07     & +0.01            & -0.06     & -0.05            & -0.03          & +0.00                \\
\texttt{num\_mods}             & +0.04     & +0.29            & -0.02     & +0.23            & -0.03     & -0.08            & -0.01          & -0.44                \\
\texttt{category\_hobby}       & -0.22     & -0.02            & +0.01     & +0.01            & +0.06     & +0.02            & -0.07          & -0.01                \\
\texttt{category\_discussion}  & +0.18     & +0.01            & -0.02     & -0.01            & -0.05     & -0.01            & +0.04          & +0.01                \\
\texttt{category\_memes}       & +0.18     & +0.03            & +0.01     & +0.00            & -0.04     & -0.05            & +0.00          & +0.01                \\
\texttt{category\_news}        & +0.03     & -0.01            & +0.01     & +0.00            & -0.01     & +0.01            & +0.00          & -0.01                \\
\texttt{category\_media}       & +0.34     & +0.02            & +0.10     & +0.00            & -0.08     & -0.04            & -0.00          & +0.01                \\
\texttt{category\_identity}    & -0.24     & -0.04            & -0.10     & -0.01            & +0.05     & +0.07            & +0.07          & -0.02                \\ \hline
\end{tabular}
}

\subsection{Covariates used for Strictness of Rule Enforcement Analyses}\label{app:covariates_enforcement}
The following table shows the covariates that were used in a logistic regression propensity score model for the rule enforcement analyses (\fig\ref{fig:removed_content_stratified_news}), along with the resulting Standardized Mean Differences (SMDs) after reweighting, and the associated propensity score model (P.S.) coefficients used in the logistic regression.

{\scriptsize
\begin{tabular}{l|rr|rr|rr|rr|}
                               & \multicolumn{8}{c|}{Amount of Removed Content Treatment Bin (Percentage of All Content)}                                           \\
\multicolumn{1}{l|}{}          & \multicolumn{2}{c|}{0\%-1\%} & \multicolumn{2}{c|}{1\%-2\%} & \multicolumn{2}{c|}{2\%-3\%} & \multicolumn{2}{c|}{>3\%}             \\ \hline
                               & SMD       & P.S. Coefficient & SMD       & P.S. Coefficient & SMD       & P.S. Coefficient & SMD            & P.S. Coefficient     \\
Covariate                      &           &                  &           &                  &           &                  &                &                      \\ \hline
\texttt{total\_items}          & -0.10     & -0.12            & +0.03     & -0.10            & +0.12     & +0.01            & +0.12          & +0.21                \\
\texttt{frac\_deleted}         & -0.58     & -0.51            & +0.02     & -0.01            & +0.30     & +0.10            & +0.67          & +0.42                \\
\texttt{mod\_workload}         & -0.08     & -0.00            & +0.04     & +0.14            & +0.15     & +0.02            & +0.04          & -0.16                \\
\texttt{num\_mods}             & -0.06     & -0.21            & -0.01     & +0.12            & -0.01     & +0.01            & +0.16          & +0.08                \\
\texttt{category\_hobby}       & +0.53     & +0.10            & -0.17     & -0.01            & -0.29     & -0.01            & -0.46          & -0.08                \\
\texttt{category\_discussion}  & -0.26     & -0.09            & +0.00     & +0.01            & +0.02     & -0.01            & +0.40          & +0.09                \\
\texttt{category\_memes}       & -0.04     & +0.08            & -0.01     & -0.00            & +0.04     & +0.01            & -0.02          & -0.08                \\
\texttt{category\_news}        & -0.11     & -0.12            & -0.06     & -0.06            & -0.00     & -0.00            & +0.28          & +0.19                \\
\texttt{category\_media}       & -0.12     & +0.03            & -0.08     & -0.01            & +0.02     & -0.00            & +0.21          & -0.02                \\
\texttt{category\_identity}    & -0.33     & +0.00            & +0.32     & +0.08            & +0.31     & +0.02            & +0.02          & -0.10                \\ \hline
\end{tabular}
}

\end{document}